\documentclass[aps,pra,reprint,amsmath,amssymb,superscriptaddress,onecolumn,longbibliography,nofootinbib,notitlepage]{revtex4-2}
\usepackage{mathtools}
\usepackage{siunitx}
\usepackage[hidelinks]{hyperref}
\usepackage{svg}
\usepackage{float}
\usepackage[braket, qm]{qcircuit}
\usepackage{bbm}
\usepackage{bm}
\usepackage{soul}
\usepackage{lipsum} 
\usepackage{tocloft}
\usepackage{titletoc}
\usepackage{graphicx}
\usepackage{adjustbox}
\usepackage{multirow}
\usepackage{lipsum} 
\usepackage[most]{tcolorbox} 
\usepackage{tikz}
\usepackage{xcolor} 
\newtcolorbox{redbox}[1]{colback=red!5!white,colframe=red!75!black,fonttitle=\bfseries,title=#1}

\newcommand{\Usense}{\mathcal{U}_{\rm sense}}
\newcommand{\Uqs}{\mathcal{U}_{\rm QS}}
\newcommand{\Uqcs}{\mathcal{U}_{\rm QCS}}

\newcommand{\Fqs}{\mathcal{F}_{\rm QS}}
\newcommand{\Fqcs}{\mathcal{F}_{\rm QCS}}

\newcommand{\Udec}{\mathcal{U}_{\rm meas}}
\newcommand{\Uenc}{\mathcal{U}_{\rm probe}}
\newcommand{\Ucoh}[1]{\mathcal{U}_{\rm coh}^{(#1)}}

\newcommand{\Ft}{\mathcal{F}^{\star}}

\usepackage{xcolor}

\begin{document}
\title{Quantum Computational Sensing}

\author{Saeed~A.~Khan}
\thanks{Equal contribution.}
\affiliation{School of Applied and Engineering Physics, Cornell University, Ithaca, NY 14853, USA}

\author{Sridhar~Prabhu}
\thanks{Equal contribution.}
\affiliation{School of Applied and Engineering Physics, Cornell University, Ithaca, NY 14853, USA}
\affiliation{Department of Physics, Cornell University, Ithaca, NY 14853, USA}

\author{Logan~G.~Wright}
\thanks{Present address: Department of Applied Physics, Yale University, New Haven, CT 06520, USA}
\affiliation{School of Applied and Engineering Physics, Cornell University, Ithaca, NY 14853, USA}

\author{Peter~L.~McMahon}
\email{pmcmahon@cornell.edu}
\affiliation{School of Applied and Engineering Physics, Cornell University, Ithaca, NY 14853, USA}
\affiliation{Kavli Institute at Cornell for Nanoscale Science, Cornell University, Ithaca, NY 14853, USA}

\newcolumntype{M}{>{\begin{varwidth}{4cm}}l<{\end{varwidth}}} 

\begin{abstract}
Quantum computing has the potential to deliver large advantages on computational tasks, but advantages for practical tasks are not yet achievable with current hardware. Quantum sensing is an entirely separate quantum technology that can provide its own kind of a quantum advantage. In this Perspective, we explain how the merger of quantum sensing with quantum computing has recently given rise to the notion of \textit{quantum computational sensing}, and a new kind of quantum advantage: a \textit{quantum computational-sensing advantage}. This advantage can be realized with far lower hardware requirements than purely computational quantum advantage. We explain how several recent proposals and experiments can be understood as quantum computational sensing, and discuss categorizations of the general architectures that quantum-computational-sensing protocols can have. We conclude with an outlook on open questions and the prospects for quantum computational sensors and quantum computational-sensing advantage.
\end{abstract}

\maketitle

\section{Introduction}
\label{sec:intro}

The past decade has seen rapid progress in the development of quantum computers, leading to the demonstration of quantum computational advantage in random-circuit sampling \cite{arute2019quantum,morvan2024phase} and Gaussian boson sampling~\cite{zhong_quantum_2020, madsen_quantum_2022}. However, realizing hopes \cite{mohseni2017commercialize} for a quantum computational advantage on tasks of practical interest, especially in machine learning, has proven elusive \cite{cerezo2022challenges, zimboras2025myths}. Meanwhile, another quantum technology---quantum sensing---has also been validated in a series of proof-of-principle demonstrations of quantum advantage~\cite{giovannetti_advances_2011,Degen_2017, lawrie_quantum_2019, zhang_entanglement-based_2024}, and has been deployed to achieve a practical advantage in, for example, gravitational-wave detection \cite{jia2024squeezing} and dark-matter searches \cite{backes2021quantum,agrawal2024stimulated,braggio2025quantum}, with quantum sensors now being engineered for practical use beyond academia \cite{stray2022quantum,bongs2023quantum,gschwendtner2024quantum,muradoglu2025quantum}. These two types of quantum advantage---quantum computational advantage and quantum sensing advantage---may at first appear completely unrelated. In this Perspective, we explain how a new kind of quantum advantage has recently emerged---a \textit{quantum computational-sensing advantage} (QCSA)---that arises from the synthesis of quantum sensing with quantum computation. This can be understood in analogy with the synthesis of classical sensing with classical computation, which is often called computational sensing \cite{vanderspiegel1996computational} or intelligent sensing \cite{ballard2021machine}.

Consider the task of identifying what kind of aircraft is passing by. A camera image or video (comprising, for example, millions of pixels) of the aircraft contains far more information than is needed to convey the answer (a few bits). The philosophy in computational sensing \cite{vanderspiegel1996computational} is to have sensors (such as cameras) be integrated with computation so that the output of a computational sensor contains only the essential information needed for the downstream task. Redundant information is discarded, and often the computation is used to extract simple features, tailored to the task, from the raw sensor data. A key motivation for classical computational sensing is to reduce the bandwidth of data that the sensor needs to output, or equivalently to increase the usefulness of the information it outputs given a constraint on bandwidth.

Quantum computational sensing (QCS) can be understood in similar terms, but instead of having a merely practical engineering constraint on the output bandwidth as in the classical case, QCS addresses the fundamental challenge that regardless of the complexity of a quantum system, a quantum measurement performed on it will reveal only a limited amount (1 bit) of classical information per qubit~\cite{nielsen2010quantum}, due to measurement-induced collapse of quantum states in quantum mechanics\footnote{For the sake of concreteness and clarity, we specialize in this paragraph to sensors comprising qubits, but the concepts we discuss generalize naturally to qudits and bosonic modes (qumodes).}. This fundamental challenge is of course faced by any application of quantum systems, and the design of any quantum algorithm must take it into account~\cite{nielsen2010quantum, wright_capacity_2019, hu_tackling_2023, gyurik2023limitations, recio2024single}. QCS involves trying to make sure that each measured bit from a quantum sensor reveals as much information \textit{relevant to the specific task} one is trying to perform as possible. This is achieved by combining a quantum sensor\footnote{Or multiple quantum sensors, potentially distributed in space, as we will discuss.} with a quantum processor that performs computations on the quantum states produced by the sensor; an advantage (QCSA) can be achieved when the computation allows more relevant information for a particular task to be obtained by a measurement of the quantum system than would have been possible without quantum processing. This then naturally leads to an advantage (QCSA; see Box~1) in sensing resources, such as sensing time: a system using a quantum computational sensor can require less time sensing to achieve the task with a given accuracy than a system using a conventional quantum sensor---which aims to extract generic rather than task-specific information. 

\begin{redbox}{Box 1: Quantum Computational-Sensing Advantage (QCSA) (intuitive definition)}
QCSA is an advantage that a quantum computational sensor can achieve over a conventional quantum sensor in extracting task-specific information from the physical world. Measurements of a quantum computational sensor can reveal more information relevant to a specific task than a generic quantum sensor would, and as a result allow a task to be achieved with a certain accuracy using less sensing time. Equivalently, QCSA can manifest as a task being achievable with a higher accuracy given a certain budget of sensing time when using a quantum computational sensor versus with a conventional quantum sensor.

\end{redbox}

This advantage can be achieved with small quantum systems, including ones that are tractable to simulate classically---sidestepping one of the major obstacles to achieving an advantage with quantum computers on purely computational tasks. QCSA is very general: it can be realized in systems that are unentangled---including ones comprising just a single qubit that is used to perform both sensing and computing\footnote{It might at first seem surprising that any useful computation can be done with a single qubit; we discuss an example in Sec.~\ref{subsec:multisense} under \emph{Protocols using learning}.}---and in entangled systems\footnote{For entangled systems, there is often a strong connection with the field of distributed quantum sensing \cite{zhang2021distributed}.}; it can be realized across sensing modalities\footnote{Such as electric-field, magnetic-field, temperature, displacement, and absorption sensing.} and hardware platforms\footnote{Such as trapped ions, neutral atoms, defects in semiconductors, quantum dots, superconducting circuits, and quantum optics, covering qubits, qudits, and qumodes (bosonic modes).}, and it can be realized for a wide range of computations---including both quantum-machine-learning algorithms \cite{biamonte2017quantum, cerezo2022challenges} and non-machine-learning-related algorithms \cite{nielsen2010quantum,martyn2021grand}.

QCSA is closely related to recent developments in gaining an advantage in learning about quantum systems using coherent protocols and/or joint quantum measurements \cite{Aharonov_2022, Huang_2022, Chen_2022, oh2024entanglement}. This line of work has highlighted how there can be exponential advantage in reading out certain properties of quantum systems, or copies of quantum systems, if the measurement protocol is carefully designed and takes advantage of entanglement. The notion of QCSA encompasses these kinds of protocols and advantages when applied to sensing scenarios, but also includes sub-exponential advantages that can arise even in unentangled systems, and that may also be important in practice. We classify various QCS schemes according to the architecture they employ, as well as the tasks they are used to perform.

The remainder of this Perspective is structured as follows: we first provide a precise definition of quantum computational sensing, and use it to explain how a diverse set of recent works can be considered as examples of this emerging paradigm. We then provide a detailed analysis of several representative examples of QCS protocols, emphasizing the wide variety of tasks that quantum computational sensing can be employed for. As we will see, QCSA is very general: it can be achieved across physical platforms from qubit-based systems to bosonic systems, for the quantum computational sensing of single-parameter, multi-parameter, and time-varying signals. By identifying common features in the construction of various QCS protocols, we introduce a taxonomy of protocols, and identify architectures that could be explored for the design of QCS protocols in the future. We discuss how quantum computational sensing could lead to practical quantum advantages using near-term quantum systems, and how QCSA relates to other types of quantum advantage. Finally, we discuss open questions and challenges in this new subfield. 

\section{Defining Quantum Computational Sensing} 

\begin{figure}[h!]
    \centering
    \includegraphics{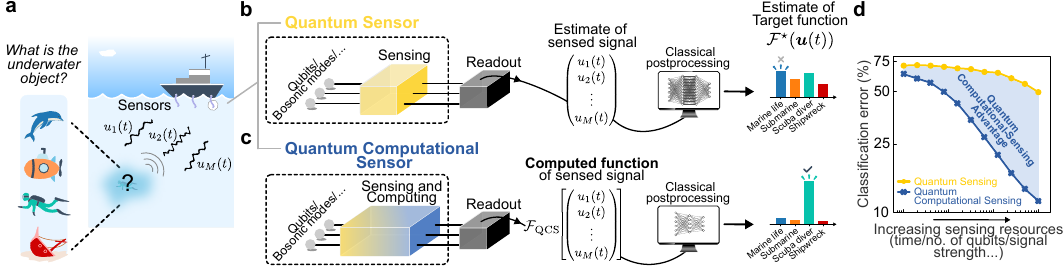}
    \caption{\textbf{Conventional Quantum Sensing versus Quantum Computational Sensing}, explained using a made-up example.\footnote{The specific setting depicted in this figure is intended to illustrate what we mean by a \emph{task} that is higher-level than raw parameter estimation, and how the paradigms of conventional quantum sensing and quantum computational sensing would typically be used to address such a classification task; our example is not intended to imply that quantum computational sensors have already been developed for the underwater-object-classification task we describe or that they would definitely provide a QCSA.} \textbf{a}, Example of a sensing task that requires extracting particular features of a physical signal $\bm{u} = (u_1(t),u_2(t),\ldots,u_M(t))$. Magnetic fields that are affected by an unidentified underwater object are sensed by a network of $M$ magnetic-field sensors. The objective of the task is to determine which of four possible objects the signal is from: a marine animal, a submarine, a scuba diver, or a shipwreck. \textbf{b}, Conventionally, a quantum sensor would be used to produce estimates of the signals $u_i(t)$. The obtained estimates of the signals would then be classically postprocessed, for example by using a classical neural network, to extract relevant features that allow determining the true identity of the underwater object. This is equivalent to computing the function $\Ft$ that yields the true class label. \textbf{c}, A quantum computational sensor combines quantum sensing with quantum computing to produce a function $\Fqcs(\bm{u}(t))$ at the output of the quantum system, rather than estimates of the raw signals $u_i(t)$. If $\Fqcs = \Ft$ (i.e., the quantum computational sensor directly outputs the target function) no further classical postprocessing would be required; otherwise, a postprocessing step (typically simpler than that in the setting of conventional quantum sensing) would be used to perform the remaining computation to output $\Ft$. \textbf{d}, A plausible (but here, again, made-up) example of quantum computational-sensing advantage (QCSA): a quantum computational sensor could potentially provide improved performance, for example a reduction in classification error, given the same sensing resources (such as the total sensing time or number of measurement samples, or number of qubits).}
    \label{fig:one}
\end{figure}

We begin with a general description of how a quantum system may be used to sense and to extract information from an unknown, multi-dimensional and potentially time-varying physical signal $\bm{u}(t)$. For example, consider the case where the physical signal $\bm{u}$ to be sensed comprises magnetic fields at different locations, and the task is that of classifying an undersea object~\cite{zhai2023detecting, ashraf2023magnetic} (see Fig.~\ref{fig:one}a). The quantum system, assumed to initially be in the state $\ket{\psi_0}$, can sense the signal $\bm{u}$ via a unitary operation $\mathcal{U}(\bm{u})$ that evolves the quantum system to the state $\ket{\psi(\bm{u})} \equiv \mathcal{U}(\bm{u})\ket{\psi_0}$. The $\bm{u}$-dependence of $\mathcal{U}$ arises from a \emph{sensing interaction}: a physical interaction between the quantum system and the unknown signal, such as the Zeeman interaction between qubits and magnetic fields $\bm{u}$~\cite{Degen_2017}. Finally, information about $\bm{u}$ is extracted from the quantum system by measurements of $\ket{\psi(\bm{u})}$. If the quantum sensor is composed of qubits, a common choice is to perform projective measurements in the computational (i.e., bitstring) basis; the use of other measurement bases can be absorbed into the definition of $\mathcal{U}$, as we will discuss. The quantum degrees of freedom or `modes' comprising such a system need not be qubits: qudits, bosonic modes and even hybrid qubit-bosonic systems can be employed for sensing applications, using analogous sensing interactions and measurement schemes.

The aim of conventional quantum sensing (QS) is to obtain a precise estimate of the unknown signal $\bm{u}$ using measurement results from a quantum sensor~\cite{giovannetti_advances_2011,Degen_2017, lawrie_quantum_2019, zhang_entanglement-based_2024} operated as we have just described. Quantum sensing protocols are typically engineered such that an unbiased estimate of the unknown signal $\bm{u}$ can be constructed directly from the measurement results. Importantly, however, measurements obtained from a quantum system are fundamentally stochastic. Unless the final state $\ket{\psi(\bm{u})}$ of the quantum system after the sensing protocol is an eigenstate of the basis in which the measurement is performed, measurement outcomes from repeated executions of the same sensing protocol can be different, exhibiting quantum sampling or projection noise. Generally, increasing the precision of an estimate of $\bm{u}$ in the presence of sampling noise requires the use of increased sensing resources. For example, one simple way to reduce the effect of sampling noise is to average the quantum sensor measurement results over multiple shots (provided the signal $\bm{u}$ remains unchanged between shots); however, this increases the total sensing time as the sensing protocol must be repeated for each shot. Other sensing resources include the number of sensing qubits, or the average photon number for bosonic systems. 

The design of quantum sensing protocols that optimize for estimation precision given constraints on sensing resources is central to the field of quantum sensing, and is an area where quantum computing has already played a role. The structure of a conventional quantum sensing protocol often takes the form $\mathcal{U} \to \Uqs = \Udec~\Usense(\bm{u})~\Uenc$~\cite{Degen_2017}. With this structure, key components of quantum sensing are the preparation of a probe state of the quantum sensor by $\Uenc$ prior to the physical sensing operation $\Usense(\bm{u})$, and preparation of an appropriate measurement basis by $\Udec$. While the simplest quantum sensing protocols can be performed even with single qubits, preparation of suitable entangled states across $M$ sensing qubits and use of entangled-basis measurements can yield an improved (Heisenberg) scaling of error as $\frac{1}{M}$, as opposed to the standard quantum limit scaling of error as $\frac{1}{\sqrt{M}}$ when the $M$ qubits are not entangled~\cite{Degen_2017}.\footnote{Similar results can be stated for quantum sensors made not from qubits but from bosonic modes, where photons are used for sensing~\cite{giovannetti2006quantum}.} Work to optimize sensing protocols to achieve the best possible parameter estimation precision given the practical limitations of the available sensing hardware has led to the recent development of variational metrology~\cite{kaubruegger2019variational, koczor2020variational, kaubruegger2021quantum, marciniak_optimal_2022, zheng2022preparation, kaubruegger_optimal_2023, maclellan2024end}, where trainable quantum circuits are used to engineer $\Uenc$ and $\Udec$ empirically. Other examples of conventional quantum sensing where quantum computing techniques are used as part of the sensing protocol include adaptive methods~\cite{direkci2024heisenberg, ma2021adaptive}, quantum-error-corrected quantum sensing~\cite{wang2022quantum, zhuang2020distributed, kessler2014quantum, arrad2014increasing, zhou2018achieving}, dynamical decoupling~\cite{titum2021optimal}, and protocols based on quantum algorithms such as quantum phase estimation~\cite{Degen_2017, motes_linear_2015} and the quantum Fourier transform~\cite{vorobyov_quantum_2021}. As such, the use of quantum computation as part of quantum sensing protocols is not new---but its use in these protocols has been restricted to helping improve the precision of estimates of the unknown parameter(s) $\bm{u}$. 

However, in many applications the precise estimation of $\bm{u}$ is not the final objective; instead, we might ultimately want to perform some higher-level task that depends on $\bm{u}$. As an example, we return to our scenario of underwater magnetic-field sensing: the ultimate objective was to determine which of the possible undersea objects the magnetic-fields signal $\bm{u}(t)$ was most likely caused by, as depicted in Fig.~\ref{fig:one}a: was it from a marine animal, a submarine, a scuba diver, or a shipwreck? Formally, this task can be cast as the computation of a function $\Ft$ of the unknown signal $\bm{u}$: we would like to compute the function $\Ft(\bm{u})$ that returns the correct class label (marine animal, submarine, scuba diver, or shipwreck) for a given input signal. In other applications, one may be interested in estimating correlation functions or other properties of $\bm{u}$, which can also be framed as estimating some function $\Ft(\bm{u})$. 

This task-specific formulation of sensing motivates an alternative approach to constructing sensing protocols: one in which the protocol is also task-specific---we call this \emph{quantum computational sensing} (QCS; see Box~2). QCS is the use of a quantum system to sense a signal $\bm{u}$ with the aim of extracting only those features of $\bm{u}$ that are relevant for a specific task. Formally, measurements of a quantum computational sensor with $\mathcal{U} \to \Uqcs$ will provide access to some function $\mathcal{F}(\bm{u})$ of the unknown physical signal, instead of the underlying signal $\bm{u}$ itself, as a conventional quantum sensor would~(see Fig.~\ref{fig:one}b). Ideally, this function will be the true target function $\Ft(\bm{u})$ for the task at hand. In the context of our example of underwater object detection, a quantum computational sensor would sense and perform a computation on the magnetic fields such that a measurement of the quantum computational sensor would, ideally, directly reveal the identity of the underwater object. It will not always be practical for a quantum computational sensor to directly output the target function $\Ft(\bm{u})$; in these cases, the quantum computational sensor can instead be engineered to output features of $\bm{u}$ that allow precise estimation of $\Ft(\bm{u})$ through classical postprocessing.

Why might one expect such an approach to be useful? After all, a conventional quantum sensor can always be used to perform the higher-level task: first the sensor would be used to obtain measurement outcomes that provide a noisy (but now classically-accessible) estimate of $\bm{u}$, following which a classical postprocessing step could be applied to compute an estimate of the function $\Ft$ that defines the task~(see Fig.~\ref{fig:one}a).

Unfortunately, this strategy is hindered by two distinct, but related factors. First, the outcomes of measurements of a quantum system are in general \textit{stochastic}, due to quantum sampling noise. As a result, the finite signal-to-noise ratio (SNR) of the measurement results will ultimately limit how well we can estimate a target function $\Ft$ by classically postprocessing the measurement results. Furthermore---and this leads to the second factor---while some types of classical postprocessing (for example, linear transformations) can preserve the SNR of the noisy measurement results, many target functions $\Ft$ would need more complex postprocessing that can degrade the SNR. For example, this difficulty is frequently encountered in the estimation of statistical moments of noisy measurement records in quantum experiments~\cite{da_silva_schemes_2010, ryan_tomography_2015}, where the SNR reduction can even be exponential in the order of the required moments. Consequently, the SNR requirements for the estimate of $\bm{u}$ in the conventional QS scheme can scale very unfavorably with the complexity of the target function $\Ft$. For a given $\Uqs$, the only way to improve the accuracy of the estimate of $\Ft$ is to increase sampling resources such as number of shots $S$, which ultimately requires increasing sensing resources such as the total sensing time. 

By directly computing functions of sensed parameters and making them accessible upon measurement, a quantum computational sensor applies a strategy to address both these limitations, as we now explain. Quantum sampling noise is fundamental, and must therefore be present in measurements of any quantum computational sensor as well; how can QCS mitigate this? The key lies in noting that in the process of computing functions of sensed parameters, the state of a quantum computational sensor must be engineered in a task-specific manner, unlike a conventional QS scheme. Being a quantum system, any such change also modifies the properties of the quantum sampling noise. This opens up the possibility of quantum computational sensors that are specifically designed to produce a given target function while simultaneously minimizing (or at least reducing) sampling noise.

Having outlined the central idea, we briefly discuss two subtle points concerning the boundary between conventional quantum sensing and quantum computational sensing. First, conventional quantum sensors often have to be carefully designed to operate in approximately linear regimes where $\Fqs(\bm{u}) = \bm{u}$; a common example is the linear operating range of Ramsey interferometers~\cite{Degen_2017}. For signals outside of this range, the conventional sensor may output functions of $\bm{u}$ more akin to quantum computational sensors. Our definition of quantum computational sensors as systems specifically designed to compute target functions of received signals excludes such conventional quantum sensors that are operated outside of the linear regime.

The second point is a related observation: the cost of task-specific design is that quantum computational sensors will (in general) not be effective \textit{conventional} quantum sensors. A simple example is a quantum computational sensor that computes $\Fqcs$ to map multiple inputs $\bm{u}$ to a single output (as in the case of classification tasks, for example). Then for a given output, the received input can not be uniquely estimated (formally, $\Fqcs$ is not an invertible map), precluding conventional sensing. A useful quantum computational sensor may therefore (by construction) be limited in the particular task it can be applied to, but for that task it would provide a QCSA.

An illuminating example is given by one of the simplest tasks for which a QCSA can be identified: computing linear combinations of multiple sensed parameters $u_i$, i.e., $\Ft = \sum_i \mathcal{W}_i u_i$. A conventional QS scheme to perform this task would first estimate the parameters independently via measurements on individual quantum sensors, and then compute the linear combination via classical postprocessing.  A QCS scheme could instead engineer the preparation of the probe state and measurement basis to directly compute the desired linear combination, and output it via measurement; it can be shown in various settings---where the sensed parameters $u_i$ are phases or displacements---that QCS can provide a more precise estimate. Even though this task requires only linear postprocessing in the conventional QS scheme, which does not degrade the SNR, a QCSA is possible due to the reduced sampling noise enabled by QCS (such quantum computational sensors have been referred to as \emph{quantum sensor networks}~\cite{eldredge2018optimal, ge2018distributed, proctor2018multiparameter, zhuang_distributed_2018,  zhang2021distributed, qian_heisenberg-scaling_2019, bringewatt2024optimal}, and are discussed in detail in Sec.~\ref{subsec:singlesense}).

However, QCS can go beyond the estimation of simple linear functions of sensed parameters. Consider the task of determining whether a weak (scalar) signal $u = \alpha$ corresponding to a displacement in a bosonic sensor is above or below a specific threshold value $\alpha_{\rm th}$, which may indicate, for example, the presence or absence of a distant target. A conventional QS protocol to estimate very weak signals is to map the received displacement to the phase of a qubit, which can then be extracted using a conventional Ramsey protocol. This can be achieved by coupling the bosonic receiver to a qubit via  $\Uenc = R_0\mathcal{D}_1R_1$ and $\Udec=\Uenc^{\dagger}$ operations; here $\mathcal{D}_j$ are bosonic displacements conditioned on the qubit state while $R_j$ are single-qubit rotations. After the protocol $\Uqs = \Udec~\Usense(\bm{u})~\Uenc$ has been applied, the qubit excitation probability is approximately $ \frac{1}{2}+(\alpha-\alpha_{\rm th})$. In other words, if the received signal is above or below the threshold, the qubit state, deviates slightly from being at the equator of the Bloch sphere in the direction of either the excited or ground state respectively. Given repeated sensing operations---as is common in conventional QS protocols---repeated measurements can be performed and the results averaged to estimate the value of $\alpha-\alpha_{\rm th}$, and the resulting threshold function can be calculated via classical postprocessing. If, on the other hand, only a single sensing operation can be performed and the qubit measured only once, very little information will be obtained: for weak $\alpha$ the excitation probability is still near $\frac{1}{2}$, and therefore in a single measurement the qubit is almost equally likely to be measured in the excited or the ground state, a manifestation of quantum sampling noise. 

In approaching this threshold task with QCS, we ask the question: can the interaction between the bosonic receiver and the readout qubit be engineered so that a single-qubit measurement directly reveals whether the signal was above or below threshold? Ref.~\cite{sinanan-singh_single-shot_2024} shows that this is indeed possible, by engineering a more sophisticated $\Uenc = R_0\prod_{j=1}^d \mathcal{D}_jR_j$ (and $\Udec=\Uenc^{\dagger}$) composed of $d$ operations. Setting $d=1$ results in the conventional QS protocol. However, increasing $d$ allows engineering an alternative protocol $\Uqcs$, called \emph{bosonic quantum signal processing}, such that the qubit excitation probability approximates the function $\frac{1}{2}+\frac{1}{2}{\rm sign}(\alpha-\alpha_{\rm th})$ (with an accuracy that increases with $d$). Engineering this highly-nonlinear relationship between qubit excitation probability and the sensed parameter $\alpha$ has an important effect: if $\alpha > \alpha_{\rm th}$ for example, the qubit under the QCS protocol is much more likely to be measured in the excited state than in the ground state when compared to the QS protocol (an analogous statement holds for $\alpha < \alpha_{\rm th}$ with the role of excited and ground states reversed). Therefore even a single measurement where the qubit is measured in the excited or ground state can now be useful: the information relevant to the thresholding task has successfully been concentrated into the single bit of information that is available in a single measurement shot. Ref.~\cite{sinanan-singh_single-shot_2024} found that the empirical single-shot classification error scales approximately as $d^{-0.82}$, so that increasingly-complex QCS protocols (larger $d$) can achieve a lower error for the same sensing time, achieving a QCSA. Further details of this QCS protocol are discussed in Sec.~\ref{subsec:singlesense}.

Achieving a QCSA relies on suitably engineering a task-specific QCS protocol (defined by $\Uqcs$). Defining QCS tasks and coming up with QCS protocols that achieve an advantage is an emerging area of research. A non-exhaustive list of recent works that fall within our definition of QCS is displayed in Fig.~\ref{fig:refs}c, organized loosely from left to right by how `complicated' the target function $\Ft$ is. We have also included several types of conventional QS, on the far left, to highlight the distinction between conventional QS and QCS: in each of the conventional QS schemes, the target function $\Ft$ is simply the identity map, $\Ft(\bm{u})=\bm{u}$. The types of functions $\Ft$ targeted using QCS can be broadly classified into two categories. First, QCS protocols can be designed for the approximation of continuous-valued functions. One of the earliest examples of QCS, shown in the left of the QCS region of the axis in Fig.~\ref{fig:refs}c, falls under this category: the development of quantum sensor networks~\cite{eldredge2018optimal, ge2018distributed, proctor2018multiparameter, zhuang_distributed_2018,  zhang2021distributed, qian_heisenberg-scaling_2019, bringewatt2024optimal} to compute linear functions of an unknown physical signal instead of estimating the signal itself. Quantum-assisted telescope arrays use the quantum Fourier transform to coherently compute the Fourier transform of the spatial correlation function of light received at different sites of the array, directly estimating the intensity distribution of the light source~\cite{khabiboulline_optical_2019}. More recently, bosonic nonlinear amplifiers have been proposed to compute arbitrary nonlinear polynomials of sensed displacements~\cite{khan_quantum_2025}. The second category is the computation of functions $\Ft$ with discrete values, with the most notable example being that of classification tasks. Supervised Learning Assisted by an Entangled Sensor Network (SLAEN)~\cite{zhuang_physical-layer_2019, xia_quantum-enhanced_2021} provides a scheme to use trainable bosonic sensors to perform linear classification tasks. Moving towards the right in Fig.~\ref{fig:refs}c, nonlinear SLAEN~\cite{liao_quantum-enhanced_2024} and bosonic quantum signal processing~\cite{sinanan-singh_single-shot_2024} are protocols for hybrid qubit-cavity sensors to approximate a thresholding function for binary classification.  Grover's algorithm can be used for the binary task of identifying the presence or absence of an oscillating signal~\cite{allen2025quantumcomputingenhancedsensing}, and Quantum signal processing can be used with single-qubit quantum sensors to perform binary classification~\cite{khan_quantum_2025}. Quantum-enhanced optical imaging protocols can coherently process light received from distinct, weak sources to directly compute functions that allow distinguishing between the sources, which can be used to perform binary classification tasks such as exoplanet detection~\cite{mokeev_enhancing_2025, gardner_quantum_2026}. Furthest to the right, various multiclass discrimination tasks have been considered, including codeword discrimination in quantum-enhanced receivers~\cite{rengaswamy_belief_2021, delaney2022demonstration, crossman2024quantum}, barcode discrimination~\cite{banchi_quantum-enhanced_2020}, quantum trajectory sensing~\cite{chin2024quantumentanglementenablessingleshot}, and classification of spatiotemporal signals~\cite{khan_quantum_2025}. In the next section, Sec.~\ref{sec:schemes}, we discuss in detail a selection of the QCS schemes shown in Fig.~\ref{fig:refs}, to highlight the variety of protocols that can enable a quantum computational-sensing advantage (QCSA), and to further explain the origins of this advantage.

\section{Schemes to achieve a Quantum Computational-Sensing Advantage}
\label{sec:schemes}

\begin{figure}[h!]
    \centering
    \includegraphics{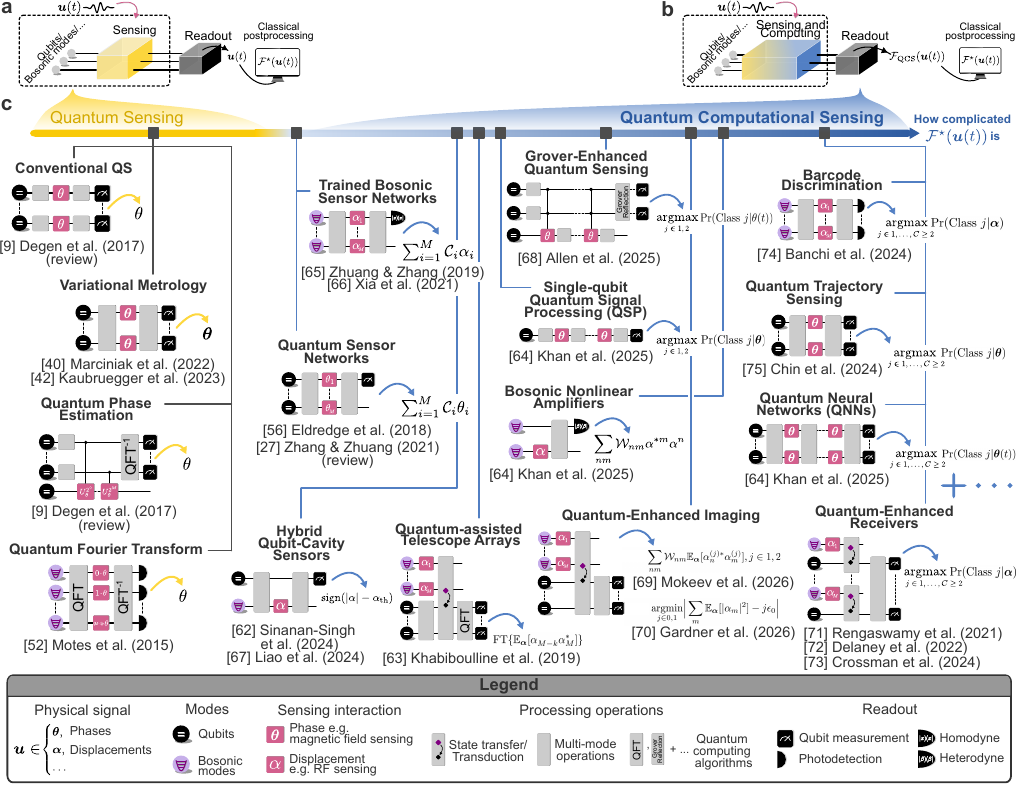}
    \caption{\textbf{Protocols for quantum computational sensing.} Computing a task-specific target function $\Ft(\bm{u}(t))$ using \textbf{a},~conventional quantum sensors, which output an estimate of the raw sensed parameters $\bm{u}(t)$, or \textbf{b},~quantum computational sensors, which output a computed function of the sensed parameters, $\Fqcs(\bm{u}(t))$. In both cases classical postprocessing of measurement results may be used to compute the final target $\Ft(\bm{u}(t))$ (although when using a quantum computational sensor, classical postprocessing is not needed if $\Fqcs$ exactly matches the target function already, i.e., $\Fqcs = \Ft$). \textbf{c},~A summary of various existing sensing protocols organized by the function $\Ft(\bm{u}(t))$ that is the target of the protocol. On the far left-hand side, conventional quantum sensing~\cite{bal_ultrasensitive_2012,loretz_radio-frequency_2013, Degen_2017} involves no computation; the sensors only output estimates of the sensed parameters $\bm{u}$ and not a function of them. Some quantum-sensing schemes (such as variational metrology~\cite{marciniak_optimal_2022, kaubruegger_optimal_2023}) perform some quantum operations, or even recognizable quantum algorithms such as quantum phase estimation~\cite{Degen_2017} or the quantum Fourier transform~\cite{motes_linear_2015}, but still with the aim of simply estimating raw parameters rather than computing functions of the sensed parameters. However, beyond the yellow--blue transition, we have systems and protocols where the output is not just an estimate of sensed parameters, but a computed function of one or more parameters---in other words, QCS protocols. The first examples of QCS here involve computing weighted linear sums (quantum sensor networks~\cite{eldredge2018optimal, zhang2021distributed} and trained bosonic sensor networks a.k.a. SLAEN~\cite{zhuang_physical-layer_2019, xia_quantum-enhanced_2021}). As we progress to the right, ever-more sophisticated computations are performed quantumly, beginning with specific nonlinear function computation tasks: threshold detection with hybrid qubit-cavity sensors~\cite{liao_quantum-enhanced_2024, sinanan-singh_single-shot_2024}), and using quantum-assisted telescope arrays~\cite{khabiboulline_optical_2019} to compute the Fourier transform of the spatial correlation function of received light (which depends on the separation $k$ between telescope sites in the array), to estimate the source light intensity distribution. We then have several binary-classification tasks: threshold detection of oscillating signals over many frequencies using Grover's search algorithm~\cite{allen2025quantumcomputingenhancedsensing}, binary classification with single-qubit quantum signal processing~\cite{khan_quantum_2025}\, and binary classification tasks such as exoplanet detection using quantum-enhanced imaging~\cite{mokeev_enhancing_2025, gardner_quantum_2026}. Next, arbitrary polynomial approximation tasks can be performed using bosonic nonlinear amplifiers~\cite{khan_quantum_2025}. Furthest to the right, we have QCS protocols for various multiclass discrimination tasks (codeword discrimination using quantum-enhanced receivers~\cite{rengaswamy_belief_2021, delaney2022demonstration, crossman2024quantum}, barcode discrimination using bosonic sensors~\cite{banchi_quantum-enhanced_2020}, quantum trajectory sensing~\cite{chin2024quantumentanglementenablessingleshot}, and classification of spatiotemporal signals using quantum neural networks~\cite{khan_quantum_2025}).}
    \label{fig:refs}
\end{figure}

The QCS protocols shown in Fig.~\ref{fig:refs} fall into one of two categories, depicted\footnote{This categorization of QCS protocols makes the assumption that sensing operations are discrete in time and that protocols can be written in the circuit model, e.g., as would happen if the sensing interaction is turned on and off in time according to a clock, or approximately if the non-sensing unitaries are much faster than the signal accumulation from sensing. It is also possible for QCS to be performed with always-on sensing where the sensing and non-sensing dynamics occur together, without a separation of timescales that allows a circuit-model representation (except via Trotterization); in this setting, protocols will all essentially have multiple sensing operations before measurement.} in Fig.~\ref{fig:taxo}: protocols that perform a single sensing operation prior to measurement, and protocols that perform multiple sensing operations prior to measurement. Within the category of using a single sensing operation before measurement, a further subdivision is natural: protocols where the only difference versus a conventional QS scheme is that the measurement basis is chosen to (ideally) be optimal for the task, and protocols where both the probe preparation and the measurement basis are optimized for the task. The category of using multiple sensing operations before measurement does not have a subdivision: protocols in this category all take the form of ``computation'' unitaries alternated with sensing unitaries, prior to a final unitary that sets the measurement basis.

Orthogonal to this categorization of protocols (by how many sensing operations occur prior to measurement) is the choice of how to design the non-sensing unitaries that appear within the protocols. Here there is another natural categorization: protocols designed \emph{with} explicit, \textit{a priori} knowledge of the target function $\Ft$; and protocols designed \emph{without} explicit knowledge of the target function $\Ft$, where the protocol and function $\Ft$ are learned from data. In the first case, we say we have \textit{analytic target functions} and can use explicit knowledge of the target function $\Ft$ for a given task in the design of the QCS protocol. Such methods apply when the target function is precisely known, e.g., if we have an explicit formula, such as $\Ft = {\rm sign}(|\alpha|-\alpha_{\rm th})$ (i.e., the target function is a threshold function for the single sensed parameter $\alpha$). In the second case, we say we have \textit{learned target functions} and do not require explicit knowledge of $\Ft$. Instead, a QCS protocol can be learned from training data by minimizing a cost function. For example, QCS tasks that can be framed as supervised-learning problems can be tackled this way. 

\subsection{Computational sensing with a single sensing operation per measurement shot}
\label{subsec:singlesense}

\begin{figure}[t]
    \centering
    \includegraphics[width=\linewidth]{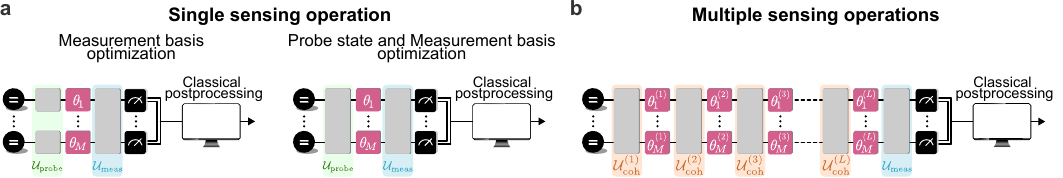}
    \caption{\textbf{Taxonomy of protocols for quantum computational sensing based on their use of either single or multiple sensing operations per measurement shot.} \textbf{a}, Structure of protocols using only a single sensing operation (per sensing mode) prior to a single measurement of the quantum sensor, described by Eq.~(\ref{eq:encdec}). Such protocols can be separated into two categories: protocols that, for a given task, optimize only the measurement basis of the quantum computational sensor (left panel), and those that optimize both the probe state and the measurement basis (right panel). \textbf{b}, Structure of protocols allowing for multiple sensing operations prior to a single measurement of the quantum sensor, described by Eq.~(\ref{eq:cohprocessing}). Such protocols rely on computing operations coherently interleaved with multiple sensing operations to perform quantum computational sensing. For concreteness, the diagrams all depict quantum computational sensors comprising qubits; the same protocol structures can be defined for other sensing and auxiliary modes (including qudits and bosonic modes), with appropriate coupling and measurement operations (both of which can also defined as continuous time, always-on processes).}
    \label{fig:taxo}
\end{figure}

Several QCS proposals consider protocols that share an important feature with conventional QS protocols: only a single sensing operation is performed before a single measurement of the quantum computational sensor is performed. Formally, the architecture of such QCS protocols is given by $\ket{\psi(\bm{u})}=\Uqcs \ket{\psi_0}$ where
\begin{align}
    \Uqcs = \Udec~\Usense(\bm{u})~\Uenc,
    \label{eq:encdec}
\end{align}
as schematically depicted in Fig.~\ref{fig:taxo}a. The operation $\Usense(\bm{u})$ describes the (typically local) sensing operation dictated by the specific physical interaction between the unknown signal $\bm{u}$ and each mode comprising the sensor; as noted before, the modes may be qubits, qudits, bosonic modes (qumodes), or combinations of these. Here we have assumed a standard simplification where $\Usense$ is taken to be a controllable interaction that can be turned on and off. In realistic scenarios, the sensing interaction may instead always be on; QCS protocols would have to be engineered to take this practical constraint into account.



\begin{redbox}{Box 2: Quantum Computational-Sensing Advantage (QCSA) (detailed definition)}
Consider the task of estimating a (possibly nonlinear) function $\Ft(\bm{u})$ of an unknown signal $\bm{u}$ to be sensed, where $\bm{u}$ may be scalar or multidimensional, and may be dependent on time.\\

\textit{Example}: Underwater object classification using magnetic field sensing~(see Fig.~\ref{fig:one}a). The signal is $\bm{u} = (B_1(t), B_2(t), \ldots, B_M(t))$---the magnitudes of the magnetic field at $M$ different underwater sensor locations. Consider the situation where the sensed magnetic fields could only have originated from one of a discrete set of underwater objects, for example a marine animal, a submarine, a scuba diver, or a shipwreck (indexed by $j=1,2,3,4$). The task is to determine the true identity of the underwater object using the sensed magnetic fields. The target function $\Ft(\bm{u})$ for this task is one that returns a probability for each object $j$ that the sensed signal originated from that object; the predicted class label is the index $j$ for which this probability is largest. \\

\textbf{Conventional Quantum Sensing (QS)}: A conventional quantum sensor is designed to obtain an estimate of the raw signal, e.g. $\Fqs(\bm{u})=\bm{u}$. A classical postprocessing operation of arbitrary complexity is then used to compute $\Ft(\bm{u})$. \\


\textit{Example:} An $M$-qubit conventional quantum sensor can be used to sense the magnetic fields $\bm{u} = (B_1(t), B_2(t), \ldots, B_M(t))$ that are incident on the sensors. Measuring the qubits in the computational basis and averaging over multiple shots $S$ yields a finite-precision estimate of the magnetic fields. These estimated time series could then be passed through a classical postprocessing operation (e.g., a classical neural network) to determine the class label that identifies the underwater object. \\

\textbf{Quantum Computational Sensing (QCS)}: A quantum sensor combined with quantum processing is used to extract information from the signal that is most relevant to ultimately estimating $\Ft(\bm{u})$. Unlike in conventional quantum sensing, measurements of a quantum computational sensor do not directly provide an estimate of $\bm{u}$, instead most generally yielding a function $\Fqcs(\bm{u})$. For example, if a quantum computational sensor for a given task can be engineered to directly compute the target function, $\Fqcs(\bm{u}) = \Ft(\bm{u})$, no further classical postprocessing is required. This is not a necessity, however, and quantum computational sensors can more generally be supplemented by some classical postprocessing to aid the computation of $\Ft(\bm{u})$. \\

\textit{Example:} The state of an $M$-qubit quantum computational sensor would evolve in a way that depends on the magnetic fields. Measurements of the quantum computational sensor would not enable estimation of the magnetic fields themselves, but instead yield a function $\Fqcs(\bm{u})$ of them. Designing a quantum computing algorithm that computes a suitable function $\Fqcs(\bm{u})$ depends on the details of the task and the quantum sensors; Section~\ref{sec:schemes} discusses various examples for different tasks. Ideally, the quantum computational sensor can be engineered such that the outcome of measuring a set of output qubits is the $j$th bitstring if and only if the underwater object is the $j$th one. In this case, measurement of the quantum computational sensor would directly predict, with 100\% accuracy using just a single measurement shot, the identity of the underwater object, with no further classical postprocessing required. Depending on the complexity of the task and the quantum sensing and computational resources available, it will not always (or even typically) be possible to achieve perfect accuracy in a single shot, even if classical postprocessing is used. We will discuss later an example of a case where perfect single-shot accuracy without classical postprocessing is achievable (Fig.~\ref{fig:schemes}c) and examples of cases where this ideal is not achieved (having less-than-100\% accuracy, needing more than one measurement shot, requiring classical postprocessing, or a combination of all three). \\

\textbf{Quantum Computational-Sensing Advantage (QCSA)}: QCSA is achieved when, given access to the same sensing resources (the same number of sensing qubits/bosonic modes/$\ldots$; the same interaction Hamiltonian between the signal and the quantum system; and the same amount of time to sense the signal), and to arbitrarily-complex classical postprocessing, a quantum computational sensor is able to provide a more accurate estimate of $\Ft(\bm{u})$ than is possible using a conventional quantum sensor.\\

\textit{Example:} A suitably designed $M$-qubit quantum computational sensor sensitive to the magnetic fields for a fixed total sensing time might be able to identify the underwater object with a lower classification error percentage than an $M$-qubit conventional quantum sensor, in which case it would exhibit a QCSA. \\

\end{redbox}



As for conventional quantum sensors, $\Uenc$ and $\Udec$ describe operations to prepare the probe state and to optimize the measurement basis respectively. Both can generally be entangling operations on the $M$ modes comprising the quantum computational sensor. Like in conventional QS, both $\Uenc$ and $\Udec$ can be engineered or numerically optimized; in contrast to conventional QS this optimization is now to enable the extraction of some specific features of $\bm{u}$, or generally to compute some function $\Ft(\bm{u})$. In variational quantum metrology (e.g., in Ref.~\cite{kaubruegger2021quantum}), where the sensing protocol often also takes the form in Eq.~(\ref{eq:encdec}), $\Uenc$ is sometimes called an \emph{encoder}, and $\Udec$ a \emph{decoder}. For the convenience of readers who like these terms for referring to probe-state preparation and measurement-basis preparation respectively, we have included them parenthetically in the discussion that follows.

\subsubsection{Measurement-basis (decoder) optimization for analytic target functions}

A subclass of protocols that use a single sensing operation per measurement are those that can achieve a QCSA by optimization of the measurement basis alone (depicted in the left panel of Fig.~\ref{fig:taxo}a); in the context of Eq.~(\ref{eq:encdec}), this involves the optimization of $\Udec$ only. Furthermore, here we only include protocols for which $\Uenc$ consists of single-mode operations, and therefore does not generate entanglement in the probe state of $M$ sensing modes.\footnote{We could have defined this category in the left panel of Fig.~\ref{fig:taxo}a so that $\Uenc$ is allowed to generate entanglement but just not to be optimized for a specific task. After all, many conventional QS schemes involve generating an entangled probe state. However, we settled on the definition used here because it allows a clearer distinction between the QCS examples we cover. The following subsection, on \textit{Probe state and measurement basis (encoder and decoder) optimization}, considers protocols where $\Uenc$ does allow the generation of entanglement.} Informally, such protocols can be considered as, following a sensing operation, identifying the `right' measurement basis to extract useful information for a specific task. 

\underline{\emph{Joint detection}}---This approach of measurement-basis optimization has been proposed and even demonstrated to provide a QCSA for communication applications. For concreteness, consider the specific example shown in Fig.~\ref{fig:schemes}a for optical communication through space. A sender transmits information by encoding it in multiple signals, typically via a modulation scheme such as binary phase-shift keying (BPSK): here each signal encodes a single bit of information, for example, in the phase ($0$ or $\pi$) of the transmitted optical signal. Multiple transmitted signals together constitute a \textit{codeword}. The receiver must sense and then `decode' the received signals to determine which of a finite set of codewords was sent; the required computation therefore naturally takes the form of a multi-class discrimination task.

A typical setting depicted in Fig.~\ref{fig:schemes}a considers the case where the receiving sensor comprises multiple optical-frequency bosonic modes, and depending on the phase of the incoming signal (which has only two possible values for BPSK) the $i$th bosonic receiver experiences a displacement $\alpha_i = \pm\alpha$. The conventional receiver strategy is to measure the individual receivers one by one, followed by classical postprocessing of the results of these measurements to determine which codeword was received. However, the nonorthogonality of coherent states means that the received states are not perfectly distinguishable, so the decoding step has an associated error. For the case of very weak signals and subsequently very small displacements $|\alpha_i|$ (relevant for long-distance communication where loss may be high), the error rate of the conventional strategy can be large.

QCS schemes for this task~\cite{guha2011quantum, guha_structured_2011} propose alternative decoding strategies by performing some form of optimized collective measurement on several, or even all, of the receiver modes. Such an optimized collective measurement can be performed in the optical domain directly, and has been experimentally demonstrated~\cite{chen2012optical}.
Alternatively, the received state can first be transduced to a qubit-based quantum processor, allowing quantum-computing operations to be used to implement more sophisticated protocols before measurement. One such protocol uses the algorithm \emph{belief propagation with quantum messages} (BPQM)~\cite{renes_belief_2017, rengaswamy_belief_2021}; the key components of the protocol are depicted in Fig.~\ref{fig:schemes}a, here for the decoding of 8 codewords, each comprising 5 bits. BPQM uses entangling unitaries, mid-circuit measurements and measurement-conditioned circuits to directly decode a minimal subset of the transmitted bits. Decoding of the remaining bits then requires only a simple XOR operation in classical postprocessing. 

The performance of BPQM can be compared against a conventional QS strategy that uses optimal measurements~\cite{helstrom1969quantum} to decode bit-by-bit, followed by an optimal classical postprocessing algorithm. The error probabilities of both schemes~\cite{rengaswamy_belief_2021} are shown in the right panel of Fig.~\ref{fig:schemes}a, showing a QCSA: the QCS scheme achieves a lower error probability than the conventional QS strategy for the same incident signal power (mean number of photons at each receiver). A more practical measurement for conventional QS would use homodyne detection of each receiver to estimate its displacement; this leads to worse performance and, when used as a baseline, results in an even larger QCSA for the BPQM scheme~\cite{delaney2022demonstration}. The decoding stage of this protocol (namely, without the receiver and transduction stages that are necessary to realize a full system) has been experimentally demonstrated~\cite{delaney2022demonstration}, using three qubits of a trapped-ion quantum processor to implement BPQM for the decoding of 4 codewords, each comprising up to 3 bits. 

\underline{\emph{Variational joint detection}}---QCS protocols relying on measurement-basis optimization can also be identified using variational methods. For the task of distinguishing codewords, Ref.~\cite{crossman2024quantum} proposes a quantum-computer-enabled receiver where measurement-basis preparation using $\Udec$ (following transduction from the optical domain to the space of computational qubits) consists of a variational quantum circuit~\cite{crossman2024quantum}. The variational circuit---which consists of interleaved CNOT operations and single-qubit rotations---can be trained to implement a decoding strategy that outperforms protocols that use optimal local measurements. While the training cost increases with problem scale (both the number of distinct codewords to be distinguished as well as the length of individual codewords), small-scale problems may be addressable with modest numbers of qubits. An initial demonstration of the algorithmic approach (but, as with BPQM above, without the receivers or transducers needed for a practical system) has been performed on a superconducting-circuit quantum processor~\cite{crossman2024quantum}.

\underline{\emph{Nonlinear amplifiers}}---In addition to discrimination tasks, QCS protocols that optimize $\Udec$ only can also be engineered to approximate arbitrary polynomial functions of the received signal $\alpha$~\cite{khan_quantum_2025}. Consider again the sensing of a weak signal that displaces a bosonic receiver by a small complex-valued displacement $\alpha$ (with real and imaginary parts defining the in-phase and quadrature components of the incident field). The conventional QS strategy is to first attempt to estimate $\alpha$ as accurately as possible, followed by classical postprocessing to compute the desired function of $\alpha$. If the target $\Ft$ is a completely arbitrary function of a small displacement $\alpha$, this strategy must incorporate a linear phase-preserving amplifier to amplify both quadratures of $\alpha$. This comes at a cost: phase-preserving amplification, due to Caves's theorem~\cite{caves_quantum_1982}, inevitably adds noise to both amplified quadratures, limiting the accuracy of the estimate of $\alpha$. Subsequent processing of this noisy estimate---required for the computation of $\Ft(\alpha)$---can degrade the signal-to-noise ratio for the estimation of many nonlinear functions.

A QCS approach~\cite{khan_quantum_2025} is to instead use a bosonic nonlinear amplifier following the sensing interaction and prior to measurement. In this case $\Udec$ takes the form of a nonlinear quantum non-demolition interaction between the sensing mode and an auxiliary bosonic mode. Such an amplifier can compute and amplify exactly the desired function $\Ft$ of the sensed displacement $\alpha$ by suitable choice of the interaction parameters in its quantum non-demolition Hamiltonian. Output from this nonlinear amplifier can have much lower noise~\cite{khan_quantum_2025, epstein_quantum_2021} than the postprocessed output of a linear phase-preserving amplifier, allowing a more accurate estimation of $\Ft$ given the same sensed signal strength. The resulting QCSA can grow with the complexity of $\Ft$, for example with the degree of a polynomial function being estimated.

\underline{\emph{Quantum-assisted telescope arrays}}---Astronomical imaging applications can also benefit from QCS protocols. For the imaging of faint stellar objects, arrays of telescope sites capable of detecting single photon excitations are used to perform optical interferometry of the collected light~\cite{gottesman_longer-baseline_2012, khabiboulline_optical_2019}. A quantity of particular interest is the intensity distribution of the stellar object, which can be related to the Fourier transform of the spatial coherence of the detected light at different telescope sites~\cite{zernike_concept_1938}. The conventional approach to estimating the intensity distribution first measures the spatial coherence of light detected at pairs of telescope sites, storing the result in classical memory, before then performing a classical Fourier transform. This involves an incoherent sum over individual spatial coherence measurements, which introduces shot noise to the final estimate of the intensity distribution.

The QCS approach instead utilizes a canonical quantum computing operation---the quantum Fourier transform---to coherently compute the Fourier transform of the spatial coherence across all telescope sites~\cite{khabiboulline_quantum-assisted_2019}, prior to measurement. Subsequent measurement of the excitation probability of the telescope sites then maps directly to the intensity distribution, with no classical postprocessing required. This approach provides a QCSA by reducing the variance in the estimate of the intensity distribution relative to the conventional approach by a factor that scales with the size of the array~\cite{khabiboulline_quantum-assisted_2019}.

\underline{\emph{Quantum-enhanced optical imaging}}---QCS protocols have also been proposed for more general tasks in optical imaging, for example classification problems. Consider the problem of identifying whether faint incoherent light received by a quantum imaging setup originated from a background source such as a star, or from a possible exoplanet of interest. Conventional approaches perform full state tomography of the received light, which can be very expensive in terms of the number of signal photons that must be received (a measure of sample complexity for this task) for an accurate reconstruction.

Ref.~\cite{mokeev_enhancing_2025} proposes interfacing the quantum imaging setup with a qubit-based quantum processor. This processor is used to perform quantum principal component analysis and quantum signal processing to sort the eigenvectors of the received state of light according to their source and without learning their precise properties. Measurements of the processor qubits can be configured to directly compute arbitrary observables on the received state. By avoiding the intermediate full reconstruction of the state of the received light, this approach avoids an unfavorable scaling of sample complexity with the dimension of the state itself, which direct detection followed by quantum state tomography suffer from. The resulting protocol is estimated to allow a sample advantage of several orders of magnitude over standard techniques for the approximation of arbitrary observables.

A similar task of incoherent sensing requires identifying a signal of interest from amongst a noisy background, in the case where only very limited information is known about the properties of the background. Ref.~\cite{gardner_quantum_2026} proposes protocols that use a quantum processor interfaced with the optical sensor to implement either weak Schur sampling or an algorithm combining density matrix exponentiation and quantum signal processing to directly test the rank, purity, and spectral gap of the unknown quantum state to detect the incoherent signal. The proposed algorithms can provide a range of sample complexity and memory advantages over direct approaches like full quantum state tomography.

\subsubsection{Measurement-basis (decoder) optimization for learned target functions}

\underline{\emph{Quantum receiver enhanced by adaptive learning}}---Learning schemes can be particularly powerful when operating conditions vary during deployment of a quantum computational sensor, for example due to noise accumulation or parameter drift over time; then, a protocol that adapts to the varying conditions may be required. The optical quantum receiver enhanced by adaptive learning (QREAL)~\cite{cui2022quantum} is an example of such a protocol optimized using a learning scheme to achieve a QCSA for distinguishing codewords. The sensing operation generates a coherent state in the sensor that is processed by the QREAL architecture, which consists of multiple processing steps or `rounds'. Each round applies a trainable quadrature displacement operation prior to a partial measurement. The complete history of measurements after a given round is used to adaptively optimize the trainable displacement operation performed in the next round using reinforcement learning. As a result, the receiver can adapt to variations in operating conditions over the course of the protocol. This improvement can be observed in practice~\cite{cui2022quantum}, where QREAL is able to outperform both conventional QS strategies (which are optimized to estimate the displacement first) and non-adaptive joint detection protocols in the presence of experimental noise and imperfections.

We have seen in this subsection that QCS protocols that rely on measurement-basis optimization alone can achieve a QCSA for some tasks (all of them centered around processing the information contained in one or more impinging bosonic coherent states), even though the unitary preparing the probe state, $\Uenc$, by construction does not generate any entanglement. However, for many tasks, creating entanglement in the probe state, in a task-specific way, will be useful---as we will discuss next.

\subsubsection{Probe state and measurement basis (encoder and decoder) optimization for analytic target functions}

We next consider QCS protocols that involve---in addition to measurement-basis optimization (by optimizing $\Udec$)---the optimization of the probe state preparation unitary $\Uenc$, which will generate task-specific entangled probe states of the modes comprising the quantum computational sensor~(Fig.~\ref{fig:taxo}a, right panel). The optimization of both the probe state and measurement basis is central to many conventional QS protocols (but for the task-independent goal of reconstructing the signal $\bm{u}$); as we will see, the same structure also forms the basis of several QCS protocols. 

\begin{figure}[h!]
    \centering
    \includegraphics[scale=0.88]{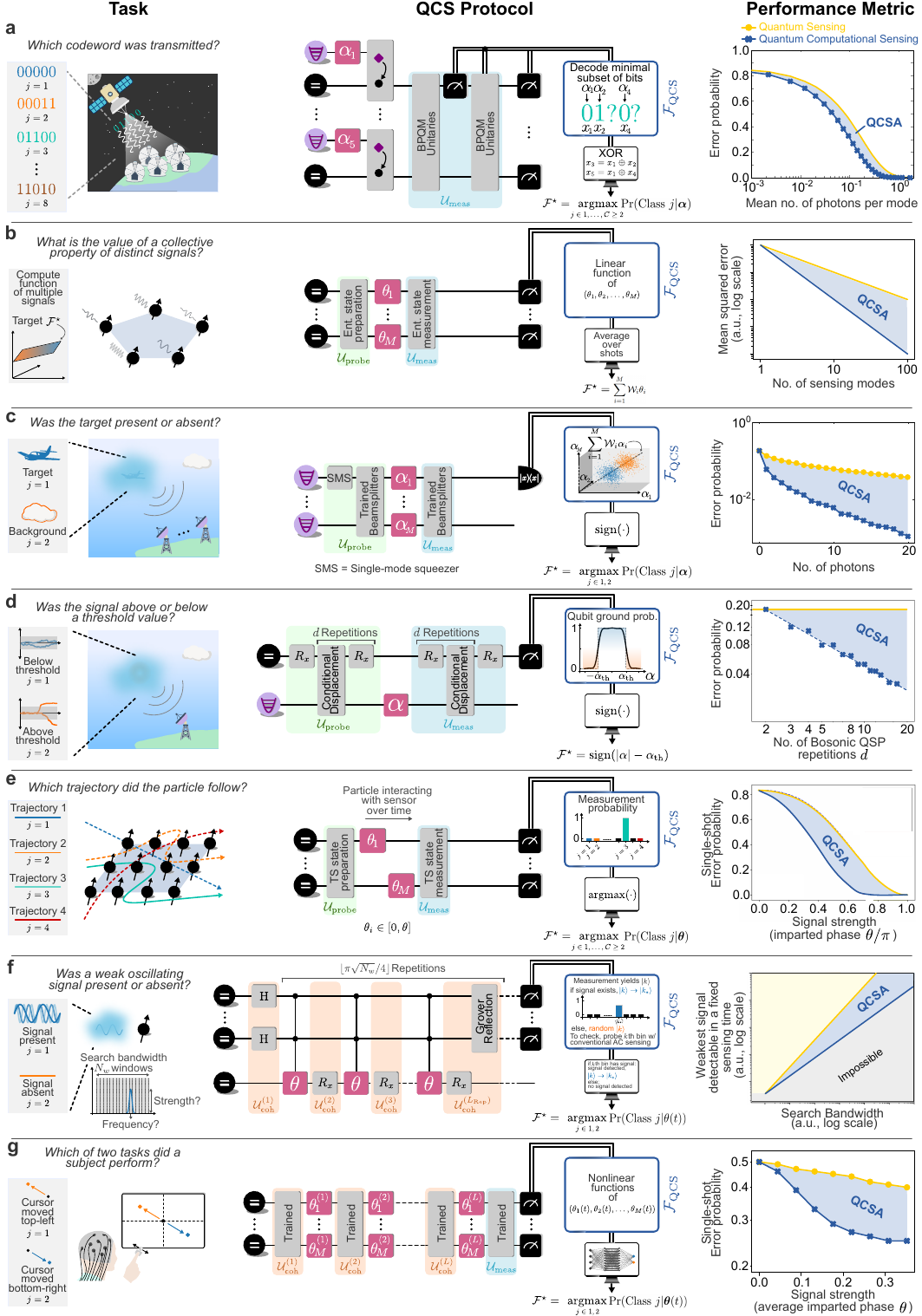}
    \caption{\textbf{Examples of QCS protocols that realize a quantum computational-sensing advantage.} \textbf{a}, Decoding transmitted codewords in long-distance communication using a quantum-enhanced receiver~\cite{rengaswamy_belief_2021,delaney2022demonstration}. \textbf{b}, Approximating functions of distinct nonlocal signals using quantum sensor networks~\cite{eldredge2018optimal, proctor2018multiparameter}. 
    \textbf{c}, Computing linear functions of received signals for target detection using trainable bosonic sensor networks~\cite{zhuang_physical-layer_2019}.
    \textbf{d}, Threshold detection of received signals using a qubit-controlled bosonic sensor via bosonic quantum signal processing~\cite{sinanan-singh_single-shot_2024}. \textbf{e}, Identifying particle trajectories using entangled qubit sensors~\cite{chin2024quantumentanglementenablessingleshot}.  \textbf{f}, Detecting the presence or absence of an oscillating signal of unknown amplitude and frequency using quantum search sensing~\cite{allen2025quantumcomputingenhancedsensing}. \textbf{g}, Classification of spatiotemporal signals received by multiqubit sensors using coherent processing with a quantum neural network~\cite{khan_quantum_2025}. }
    \label{fig:schemes}
\end{figure}
\clearpage

\underline{\emph{Quantum sensor networks}}---An illuminating example in this category is also one of the earliest QCS protocols, and one we have referred to already: quantum sensor networks (QSNs)~\cite{ge2018distributed, eldredge2018optimal, proctor2018multiparameter,zhuang_distributed_2018,qian_heisenberg-scaling_2019,  guo2020distributed, xia_demonstration_2020, rubio2020quantum, qian2021optimal, bringewatt2021protocols, liu_distributed_2021, zhang2021distributed, xia_entanglement-enhanced_2023, bringewatt2024optimal, van2024utilizing}, depicted in Fig.~\ref{fig:schemes}b. Qubit-based QSNs consider $M$-qubit sensors, where the $i$th qubit is sensitive to a physical signal (say a magnetic or electric field) that leads to it accumulating a phase $\theta_i$. The task is to estimate an arbitrary linear combination of the phases, $\Ft = \sum_i \mathcal{W}_i \theta_i$. Since the phases sensed by each qubit are different, the conventional QS strategy would first produce estimates of each of the phases $\theta_i$ and then compute $\Ft$ in classical postprocessing. This approach does not use any entanglement between the qubits. Instead, QSNs use an appropriate $\Uenc$ to entangle the $M$ qubits prior to the sensing operations, and $\Udec$ to enable a specific measurement of the sensor network, such that the measurement directly estimates the weighted linear combination of interest. A simple example to gain intuition for how such QSNs can achieve a QCSA is that of the estimation of an unweighted sum ($\mathcal{W}_i = 1~\forall~i$) of phases $\theta_i$. If the phases $\theta_i$ were all equal across the $M$ qubits, this task reduces to conventional parameter estimation. For this special case ($\mathcal{W}_i = 1$ and $\theta_i=\theta~\forall~i$), an optimal conventional approach is to have $\Uenc$ prepare a GHZ probe state, and use $\Udec$ to perform a measurement in the GHZ basis to directly measure $M\sum_i\theta = M\theta$. The coherent summing of $\theta$ enhances the signal by a factor of $M$, which gives Heisenberg scaling in error~\cite{Degen_2017}. For distinct $\theta_i$, this same protocol directly estimates the linear function $\Ft = \sum_i \theta_i$, achieving a generalized Heisenberg limit for the approximation of linear functions~\cite{eldredge2018optimal}, a simple example of QCSA.

QSNs are able to estimate not just unweighted sums, but arbitrary linear functions of an unknown multiparameter signal. QSNs measuring arbitrary weighted linear combinations can be used in adaptive schemes to optimally approximate arbitrary nonlinear functions~\cite{qian_heisenberg-scaling_2019}, as well as multiple functions~\cite{bringewatt2021protocols}. Analogous ideas can also be realized in sensor networks composed of bosonic modes~\cite{ge2018distributed, bringewatt2024optimal}, and quantum sensor networks have been experimentally demonstrated across various hardware platforms~\cite{guo2020distributed, xia_quantum-enhanced_2021, malia_distributed_2022, xia_entanglement-enhanced_2023}. While QSNs can be used to approximate arbitrary functions using an adaptive protocol, individual measurements of QSNs only provide estimates of weighted linear combinations of sensed parameters.

\underline{\emph{Bosonic quantum signal processing}}---Other QCS protocols have been proposed where measurement yields a more complex function of the sensed parameters. For example, consider the binary classification task of determining whether an incoming electromagnetic signal has magnitude above or below a predetermined threshold, as depicted in Fig.~\ref{fig:schemes}b. This signal is received by a single-mode bosonic receiver, which undergoes a displacement $\alpha$; the objective of the task is to determine whether $|\alpha|$ is above or below a threshold value $\alpha_{\rm th}$. Formally, this requires estimating the function $\Ft = {\rm sign}(|\alpha|-\alpha_{\rm th})$, a nonlinear function of the unknown parameter $\alpha$.

A conventional QS protocol would first estimate $\alpha$. For weak displacements, a cat-state sensing protocol can offer excellent sensitivity: the receiver is coupled to a single qubit, measurement of which can be used to extract information about the incoming signal~\cite{sinanan-singh_single-shot_2024}. In this setting the qubit provides a resource to engineer $\Uenc$, allowing the generation of useful probe states such as cat states prior to the sensing operation. In the cat-state sensing strategy, the qubit excitation probability after the sensing operation can be used to estimate $\alpha$, with a sensitivity and range that depends on the size of the cat state. This strategy is optimal in estimating $\alpha$, as long as it is known to lie within a certain range. Classical postprocessing can then used to estimate the required function $\Ft$ and perform the binary classification task.

Ref.~\cite{sinanan-singh_single-shot_2024} proposes a QCS protocol that is able to engineer the qubit excitation probability to have a parameterized, nonlinear dependence on the sensed displacement $\alpha$. This protocol relies on bosonic quantum signal processing (BQSP), a generalization of the qubit-based quantum signal processing (QSP)~\cite{martyn2021grand} algorithm to hybrid qubit-bosonic systems. Using $\Uenc$ composed of single-qubit rotations interleaved with qubit-state-conditioned displacements of the bosonic mode, and with $\Udec = \Uenc^{\dagger}$, BQSP enables the qubit excitation probability to have a functional dependence on polynomials of the displacement $\alpha$. The complexity of the function depends on the number $d$ of conditional-displacement operations included in $\Uenc$ (see Fig.~\ref{fig:schemes}b, center panel), as well as the values of parameters characterizing these operations. Ref.~\cite{sinanan-singh_single-shot_2024} used numerical optimization to find the parameters such that measuring the qubit at the end of the protocol gives an approximation of $\Ft$. In the special case of $d=1$, the protocol reduces to the conventional cat-state QS protocol that simply outputs an estimate of $\alpha$, while $d>1$ corresponds to QCS, allowing the quantum computational sensor to construct an increasingly-faithful approximation to $\Ft$. The error in performing the threshold task scales approximately as $d^{-0.82}$, indicating an increasing QCSA as a function of $d$ (see Fig.~\ref{fig:schemes}b, rightmost panel).

\underline{\emph{Quantum trajectory sensing}}---QCS protocols requiring probe-state and measurement-basis optimization have also been proposed for tasks beyond binary classification, such as quantum trajectory sensing (QTS)~\cite{chin2024quantumentanglementenablessingleshot}, depicted in Fig.~\ref{fig:schemes}c. Consider a particle traversing through a grid of $M$ qubits. By detecting which set of qubits accrue a phase $\theta$ from the weak interaction with this particle, the path or trajectory of the particle can be inferred. The objective of the QTS problem is to determine which of a finite set of ($>2$) trajectories the particle followed---a multiclass discrimination task. Related work on identifying which of a set of qubits (if any) has interacted with a signal has also been considered in the context of discrete-outcome sensor networks~\cite{hillery_discrete_2023, ali_discrete-outcome_2024}.

A conventional QS strategy would attempt to estimate the phase imparted on each qubit individually. However, since the result of quantum measurement exhibits quantum sampling noise, it is not possible, using just a single measurement shot, to perfectly distinguish qubits that have accrued a non-zero but small phase from those that have not. In general multiple measurements of the qubits will have to be performed to be able to accurately infer the particle trajectory in classical postprocessing. However, multiple measurements are only meaningful to perform if, for the duration that the experiment is run, the sensor array has a sequence of particles impinging on it that all follow the same trajectory.

The QCS protocol in Ref.~\cite{chin2024quantumentanglementenablessingleshot} instead extracts information directly about the trajectory by, in part, preparing a specific entangled state across the $M$ qubits using $\Uenc$. As shown in the right panel of Fig.~\ref{fig:schemes}c, the QCS approach outperforms the conventional QS strategy in terms of the probability of correctly determining the trajectory using a single measurement shot~\cite{chin2024quantumentanglementenablessingleshot}. Impressively, for a large enough phase $\theta$ imparted by the particle on each qubit that it traverses, the QCS protocol can even achieve \emph{perfect} classification accuracy using a single shot, whereas the conventional QS strategy would have a probability of error of up to approximately 20\%.

\subsubsection{Probe-state and measurement-basis (encoder and decoder) optimization for learned target functions}

\underline{\emph{Supervised learning assisted by an entangled sensor network}}---The optimization of both the probe-state and measurement basis-preparation steps for QCS protocols can also be performed by learning from data. Supervised Learning Assisted by an Entangled sensor Network (SLAEN)~\cite{zhuang_physical-layer_2019} was proposed for the setting of sensing electromagnetic signals that cause displacements to the sensor modes (Fig.~\ref{fig:schemes}d). The task is to determine whether or not the received signals were reflected off an airborne target such as an aircraft, and ultimately to perform the binary classification task of outputting whether or not the target was present. A network of bosonic modes can be used to sense the incoming signals; each mode is assumed to undergo a displacement $\alpha_i$, where $i$ indexes the modes. The classification task can be performed by determining the sign of a specific weighted linear combination of the sensed signals~\cite{zhuang_physical-layer_2019}, namely $\Ft = {\rm sign}( \sum_i \mathcal{W}_i \alpha_i + b)$.

The conventional QS strategy would be to obtain an estimate of each $\alpha_i$ as accurately as possible, which can be done by measuring the bosonic modes individually. The class label (\emph{target present} or \emph{target absent}) can then be determined using classical postprocessing of the measurement results. The sensitivity of the individual displacement measurements in the conventional QS strategy can be improved by having $\Uenc$ prepare each mode in a single-mode squeezed state, prior to the sensing (displacement) operation. In the limit of infinite squeezing, such a scheme can yield arbitrary precision. However, with practically achievable levels of squeezing, estimates of small displacements $\bm{\alpha}$ will be noisy, limiting the accuracy with which the discrimination task can be performed.

The SLAEN approach provides a QCS protocol to directly measure the desired linear combination, improving the accuracy of its estimation. This is done using a more general $\Uenc$ than in the conventional QS strategy: in SLAEN, $\Uenc$ is composed of squeezing of a single mode and beam splitters across the $M$ modes comprising the network (see Fig.~\ref{fig:schemes}d, center panel). These components can be used to prepare general multimode Gaussian entangled states prior to the sensing operation~\cite{braunstein_squeezing_2005}. The beam splitter operations comprising $\Uenc$ are variational, with the splitting ratios trained using knowledge of the weights $\mathcal{W}_i$ that characterize the target linear function $\Ft$. Following the sensing operation, measurement-basis optimization using $\Udec$ involves performing the inverse of the beam splitter operations applied during $\Uenc$, followed by measurement of a specific single mode via homodyne detection. The implemented protocol ensures that measurements of this single mode directly estimate $\sum_i \mathcal{W}_i \alpha_i + b$; the remaining ${\rm sign}(\cdot)$ operation is performed in classical postprocessing to obtain $\Ft$.

As with QSNs, since the desired target function $\Ft$ is linear in the sensed parameters, the SLAEN framework admits a simple interpretation. The probe state preparation operation $\Uenc$ performs squeezing to reduce noise in a particular `supermode' of the $M$ sensors, which is aligned with the direction set by the target linear combination $\Ft$. The measurement unitary then enables measurement in this basis. For a fixed photon-number budget in the bosonic sensor network, the SLAEN QCS protocol~\cite{zhuang_physical-layer_2019} can provide a reduced classification error versus a conventional QS approach (see Fig.~\ref{fig:schemes}d, right panel).

\underline{\emph{Nonlinear SLAEN}}---The use of purely Gaussian operations in $\Uenc$ and $\Udec$, as in SLAEN, constrains the class of functions $\Ft$ that can be computed. This limitation can be addressed by extending the protocol to include non-Gaussian operations, via a hybrid receiver where the sensing bosonic mode is coupled to a single qubit~\cite{Liao_2024, khan_quantum_2025}. Similar to the BQSP protocol, incorporating qubit-state-conditioned bosonic displacement operations as part of $\Uenc$ and $\Udec$ allows the QCS protocol to use non-Gaussian operations. By training these now more complex $\Uenc$ and $\Udec$ operations using supervised learning for binary classification tasks, a reduced classification error versus that of the conventional QS approach of direct measurement of the sensing bosonic mode can be achieved: Ref.~\cite{Liao_2024} showed this for the task of detecting if a displacement exceeded a threshold, and Ref.~\cite{khan_quantum_2025} for the task of classifying complex-valued displacements separated by nonlinear decision boundaries.

\underline{\emph{Pattern recognition}}---For tasks beyond binary classification, an example of the design of QCS protocols using learning schemes is that of barcode discrimination and pattern recognition~\cite{banchi_quantum-enhanced_2020, ortolano2023quantum}. Barcode discrimination here is the problem of distinguishing digital images that comprise only black or white pixels. Cast as a sensing problem, one can consider a set of bosonic modes, each sensitive to one pixel of the barcode. The objective of the task is to discriminate between a set of pixel patterns or `barcodes' using measurements of the bosonic modes following the sensing operation. A conventional QS strategy would attempt to faithfully reconstruct the signal (i.e., the pixels of the pattern), and then perform the classification in postprocessing. The QCS protocol instead aims to classify directly the sensed pattern. The protocol starts with $\Uenc$ that entangles pairs of bosonic modes via two-mode squeezing, and then performs an optimized two-mode POVM, described by $\Udec$, after the sensing interaction. Ref.~\cite{banchi_quantum-enhanced_2020} shows that the QCS strategy with optimized $\Udec$ and with two-mode squeezed input states can obtain a QCSA over a conventional QS strategy.

\subsection{Computational sensing with multiple sensing operations per measurement shot}
\label{subsec:multisense}

We have thus far described QCS protocols with the structure in Eq.~(\ref{eq:encdec}) where a single execution of the protocol---terminating in measurement---involves only a single instance of the sensing interaction $\Usense$. As we will see, a quantum computational sensor will often be able to compute more sophisticated functions $\Ft$ if the protocol structure is generalized to allow multiple sensing interactions, interleaved with coherent processing, before a single measurement is performed. Formally, we can describe a QCS protocol $\ket{\psi(\bm{u})}=\Uqcs \ket{\psi_0}$ with this structure as having the form:
\begin{align}
    \Uqcs = \Udec\left[ \prod_{l=1}^L \Usense(\bm{u}^{(l)})~\Ucoh{l} \right],
    \label{eq:cohprocessing}
\end{align}
where we have introduced $L$ `coherent processing' operations $\Ucoh{l}$, each of which can vary with the layer index $l$. If $L=1$, this protocol reduces to the structure we have considered thus far, described by Eq.~(\ref{eq:encdec}): a single sensing interaction is bookended by probe-state preparation (with $\Ucoh{1} \to \Uenc$) and measurement-basis preparation $\Udec$. 

For $L> 1$, however, this quantum computational sensor is able to sense an unknown signal $\bm{u}$ multiple ($L$) times and perform $L$ coherent operations prior to measurement (see Fig.~\ref{fig:taxo}b); we will therefore also refer to protocols with this structure as coherent-processing QCS protocols. Note that the argument of $\Usense$ is now also marked by the layer index $l$, emphasizing that the unknown input signal can generally vary with $l$. This scenario arises naturally in the sensing of time-varying signals, where the sensor receives a different snapshot of the signal over time, $\bm{u}^{(l)} \equiv \bm{u}(t_l)$. However, coherent processing protocols are not limited to time-varying signals only; they can also describe the processing of static signals repeated over time. While QCS schemes that rely on a single sensing operation can harness entanglement, a quantum computational sensor described by Eq.~(\ref{eq:cohprocessing}) can also sense and process signals over time; this provides an additional pathway to achieving QCSA by exploiting temporal coherence.

\subsubsection{Protocol for an analytic target function}
\label{sec:Grover}

\underline{\emph{Grover-enhanced search sensing}}---An example of a coherent processing protocol has recently been proven to be the optimal protocol for the task of detecting the presence or absence of an oscillating, time-varying signal~\cite{allen2025quantumcomputingenhancedsensing}. Consider the situation where a weak, time-varying field---whose magnitude, frequency, and phase are all unknown---might or might not be present. A qubit can sense this field (if it is present), as depicted in Fig.~\ref{fig:schemes}e. The task is to determine whether this oscillating signal with otherwise unknown properties was present or not.

As both the frequency and magnitude of the signal are unknown, one must search for the signal in a specific bandwidth $|\Delta\omega|$, and for signals above a certain minimum strength $B_{\rm min}$. A conventional approach to doing so could proceed as follows: first, a specified search bandwidth $|\Delta\omega|$ is divided into a number of windows or `bins' $N_\textrm{w}$. If the signal exists, its frequency is given by $\omega_{k_*}$, and is assumed to fall into one of these bins. We can then determine whether the incident signal exists in any one of these bins using conventional frequency-estimation techniques such as repeated $X$-basis qubit rotations followed by a Ramsey measurement. However, such a protocol must be repeated for all bins across the search bandwidth; as a result, the total sensing time required will scale linearly with the search bandwidth $|\Delta\omega|$. For a given total sensing time, this imposes a tradeoff on the task performance: a weaker signal requires longer to detect, and so a smaller bandwidth can be searched, while increasing the bandwidth reduces the sensitivity (minimum detectable signal) of the protocol. The tradeoff is depicted in the rightmost panel of Fig.~\ref{fig:schemes}f. The use of entangled sensors enables a Heisenberg scaling of sensing time, and determines the lower bound for the performance of conventional QS approaches to this task, given by the solid yellow line. 

The QCS approach of Ref.~\cite{allen2025quantumcomputingenhancedsensing} tackles the AC-field-detection task by, in part, framing it as a search problem: identifying a specific element (here a specific frequency bin) among a list (the bins comprising the search bandwidth). By setting up the search problem to be based on a phase-flip oracle, the QCS protocol can take advantage of Grover's algorithm to achieve a quadratic speedup in the search~\cite{grover_fast_1996}.

A key technical aspect of this QCS protocol is how to engineer the desired phase-flip oracle from a standard sensing interaction. The oracle, and ultimately the Grover search, acts on the sensing qubit and a set of computing qubits that are used to represent the frequency bins in the search space. Depending on the frequency of the sensed signal, the oracle needs to mark the state $\ket{k_*}$ of the computing qubits that corresponds to the frequency bin $k_*$ that the signal falls in. The oracle needs to be constructed to (i) depend only on the frequency and amplitude of the signal, and not its phase; (ii) act on frequency bins rather than individual frequencies; (iii) have a digitized response: it should flip the phase of the marked bin, and leave all other bins unchanged. The first two criteria are satisfied by appropriate choice of a time-dependent Hamiltonian coupling the sensing qubit to the computing qubits. The third criterion is achieved by using another quantum algorithm as a subroutine: quantum signal processing (QSP)~\cite{martyn2021grand} is used to realize a threshold function that digitizes the accrued phases. The sensing interaction $\Usense$ appears multiple times within the QSP subroutine that is part of the search oracle construction. The oracle itself is executed $O(\sqrt{N_\textrm{w}})$ times (where $N$ is the number of bins to search)---so the sensing interaction occurs many ($O(\sqrt{N_\textrm{w}})$) times, interleaved with coherent processing, before a measurement takes place.

Measuring the computational qubits in the computational basis yields the marked state $\ket{k_*}$ with high probability ($\geq \frac{1}{2}$) if a signal was detected, or a random state if no signal was detected. A final verification step can then be carried out to distinguish between these two outcomes, for example by using a conventional approach (which is efficient, now that only a single bin must be checked) to validate the results of the quantum search.

To compare the QCS protocol vs all possible conventional QS approaches, we can ask: given a fixed sensing time, number of sensing modes $M$, and a search bandwidth $|\Delta\omega|$, what is the weakest signal strength $B_{\rm min}$ that can be detected by the QCS protocol or by a conventional QS protocol? In both cases, if a larger bandwidth has to be searched in a fixed sensing time, the signal detection threshold $B_{\rm min}$ must go up. For a Heisenberg-limited conventional QS protocol, Ref.~\cite{allen2025quantumcomputingenhancedsensing} shows that $B_{\rm min} \propto |\Delta\omega|^{1/2}$. For the QCS protocol, $B_{\rm min} \propto |\Delta\omega|^{1/3}$ instead, so that the signal detection threshold exhibits a polynomially-slower increase with the search bandwidth. The upshot is that the QCS approach can detect weaker signals $B<B_{\rm min}$ for a fixed search bandwidth (see Fig.~\ref{fig:schemes}e, rightmost panel). Furthermore, Ref.~\cite{allen2025quantumcomputingenhancedsensing} shows that the QCS protocol for this field-detection task is asymptotically optimal (in the number of sensing qubits used) ~\cite{allen2025quantumcomputingenhancedsensing}, saturating the lower bound on the weakest AC field magnitude $B_{\rm min}$ detectable given a specific search bandwidth $|\Delta\omega|$. This example illustrates how standard quantum algorithms---in this case, Grover's algorithm and quantum signal processing---can be used to gain a provable scaling advantage in a sensing task.

\subsubsection{Protocols for learned target functions}

\underline{\emph{Quantum signal processing}}---Learning schemes can be used to optimize processing operations $\Ucoh{l}$ in coherent processing protocols when the target function $\Ft$ is not \textit{a priori} known. Consider the task of classifying, into one of two classes, the phase $\theta$ sensed by a single qubit, where the two classes are defined implicitly by a training dataset (pairs of phase and its corresponding class label)~\cite{khan_quantum_2025}. This binary classification task can be framed as the computation of a binary, nonlinear function $\Ft(u=\theta)$ that returns 0 if $\theta$ falls in the first class, and 1 if $\theta$ falls in the second class.

The conventional QS strategy would be to perform Ramsey interferometry with the single qubit to estimate the phase $\theta$ and then to perform classical postprocessing to map the estimate of $\theta$ to a class label. As a single measurement of the qubit yields only a binary outcome, multiple measurements or shots $S$ are required to obtain a precise estimate of $\theta$; the parameter $\theta$ is assumed to not change from shot to shot. The total number $N$ of sensing operations $\Usense$ required in executing $S$ shots is $N=S$; the number $N$ quantifies how much sensing time is used to perform the task.  

A coherent-processing QCS protocol~\cite{khan_quantum_2025} can achieve an advantage over conventional QS by directly approximating the binary classification function $\Ft$: for a given budget of total sensing operations $N$, one can obtain a lower classification error. Computing the function $\Ft$ relies on performing $L>1$ sensing operations before measurement; the sensing operations are interleaved with computing unitaries $\Ucoh{l}$. For fixed $N$, fewer measurement shots $S$ are allowed as $L$ increases, since $N=L\times S$, which at first might appear to be a disadvantage: per measurement shot (and hence classical bit of information extracted from the quantum system), we are now using more sensing time. However, instead of outputting an approximation of $\theta$, the system outputs an approximation of $\Ft$. It turns out that this tradeoff of using more sensing operations per shot to obtain more task-relevant information, even though the total number of measurement shots must then be reduced by more than $L\times$ for there to be a QCSA, is worthwhile.

The choice of $\Ucoh{l}$ is motivated by the quantum-signal-processing (QSP)~\cite{martyn2021grand} algorithm, which can be used to produce nonlinear functions of the sensed phase $\theta$: assuming $\theta$ couples to the sensing qubit in the $Z$ basis, $\Ucoh{l}$ are qubit rotations about an axis orthogonal to the sensing axis, for example the $X$ axis. The parameters characterizing $\Ucoh{l}$ are chosen through optimization. In this setting where the explicit form for the target function $\Ft$ is not known, the parameters of $\Ucoh{l}$ are learned using a labeled training dataset. Ref.~\cite{khan_quantum_2025} reported a QCSA where the probability of error was as much as 25 percentage points lower for QCS than for conventional QS, given an equal number of sensing operations $N$ for both. 

\underline{\emph{Quantum neural networks}}---Coherent-processing QCS protocols can also be optimized via learning schemes for tasks where a time-varying signal is sensed, and hence each subsequent sensing operation in general senses a different parameter value. An example task analyzed in Ref.~\cite{khan_quantum_2025} is the classification of time-varying magnetic field signals at $M$ different spatial locations, sensed by $M$ qubits (Fig.~\ref{fig:schemes}f, left panel). The dataset studied came from a experimental measurements of magnetic fields from neural activity in human subjects in the process of performing physical movements~\cite{yeom2023magnetoencephalography}; the task was to perform binary classification of which of two movements the subject executed.

A conventional QS strategy would perform Ramsey interferometry $N$ times sequentially for each of the $M$ qubits, yielding estimates of the $M$ time-varying signals $u_i(t)$ ($i\in\{1,\ldots,M\})$. These estimated signals could then be processed by a classical neural network to obtain a classification prediction.

The QCS protocol instead makes use of entanglement and temporal coherence in a quantum computational sensor to perform $N$ sensing operations and to coherently process them using $L=N$ processing operations $\Ucoh{l}$, before performing a single measurement of each of the qubits. The processing operations are trainable, arbitrary $M$-qubit gates; the overall protocol therefore takes on the form of a quantum neural network (QNN)~\cite{farhi2018classificationquantumneuralnetworks,mitarai2018quantum} (Fig.~\ref{fig:schemes}f, center panel). The results of the measurement of the qubits (a bitstring of length $M$) are processed by a classical neural network to obtain a classification prediction. The QCS protocol achieves an advantage over conventional QS in classification error probability by as much as 16 percentage points.

\subsection{Classification of protocols by coherence in time and entanglement across space}

\begin{figure}[t!]
    \centering
    \includegraphics[width=\linewidth]{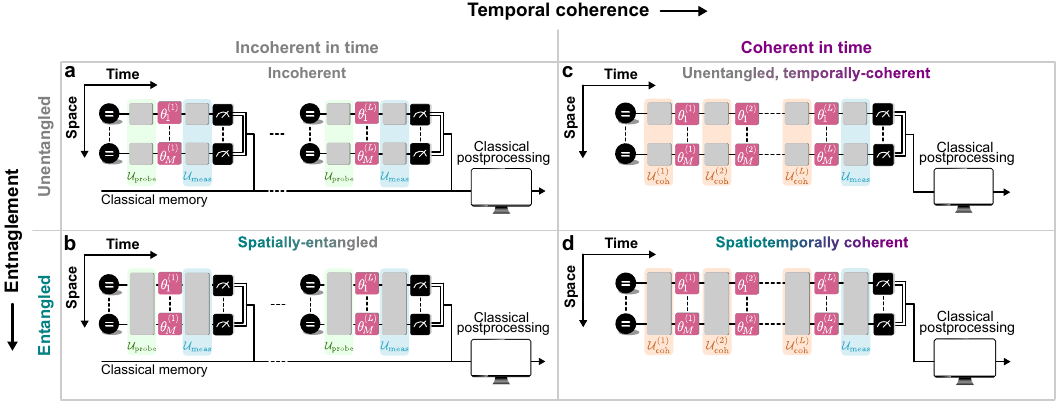}
    \caption{\textbf{Taxonomy of quantum-sensing and quantum-computational-sensing protocols based on their coherence in time and entanglement across space.}\footnote{We assume for pedagogical convenience here that the qubits/qudits/qumodes of the system are spread out in space, but other possibilities---such as multiplexing in frequency---are also compatible with the classification we give.} \textbf{a}, \emph{Incoherent}: Structure of a protocol featuring incoherent processing in time and no entanglement. \textbf{b}, \emph{Spatially-entangled}: Structure of a protocol featuring incoherent processing in time and entanglement across space. \textbf{c}, \emph{Unentangled, temporally coherent}: Structure of a protocol featuring coherent processing in time and no entanglement. \textbf{d}, \emph{Spatiotemporally coherent}: Structure of a protocol featuring coherent processing in time and entanglement across space. For concreteness, the protocol structures are all depicted for quantum computational sensors comprising qubits; the same circuit architecture can be used for qudits or qumodes, with appropriate coupling and measurement operations.}
    \label{fig:architecture}
\end{figure}

The protocols we have surveyed above in many cases differ in their use of coherence, suggesting another possible classification of protocols. Fig.~\ref{fig:architecture} shows a taxonomy of QS and QCS protocols based on whether they are incoherent or coherent in time (left vs right columns), and whether they are unentangled or entangled across space (top vs bottom rows). The most basic conventional QS protocols---for example, Ramsey interferometry performed independently on each qubit---fall in the top-left quadrant (incoherent in time; unentangled across space). Conventional QS protocols achieving Heisenberg scaling, such as interferometry using GHZ states, fall in the bottom-left quadrant (incoherent in time; entangled across space), as do the majority of examples of QCS protocols appearing in Table~\ref{table:examples_qcsa}, where there is only a single sensing operation prior to measurement. However, we have seen in our discussion of single-qubit quantum signal processing~\cite{khan_quantum_2025} that allowing a protocol to coherently compute on sensed parameters multiple times before measurement, even when there is no entanglement---corresponding to the top-right quadrant (coherent in time; unentangled across space)---can enable an advantage for a quantum computational sensor. Finally, protocols such as Grover-based signal detection~\cite{allen2025quantumcomputingenhancedsensing} and quantum neural networks~\cite{khan_quantum_2025} fall in the bottom-right quadrant (coherent in time; entangled across space). Ref.~\cite{khan_quantum_2025} compared the performance of protocols restricted to each of the four quadrants and reported that, for the task considered, QCS protocol taking advantage of both coherent processing in time and entanglement across space delivered the best results.

\section{Designing new protocols for quantum computational sensing}

The examples we have discussed are a representative subset of the QCS protocols that have been invented to date; through them we have explored some of the types of tasks that QCS can address, and identified broad categorizations of QCS protocols, such as single versus multiple sensing operations per measurement, and whether a protocol is designed for an analytic description of the target function $\Ft$, or optimized via learning schemes. Table~\ref{table:examples_qcsa} summarizes many of the existing QCS protocols by the quantum algorithm at the core of the computation done in the quantum domain, the number of sensing operations per measurement, and whether the target function is analytic or learned. It also features examples from the literature on learning of quantum systems, such as Ref.~\cite{Huang_2022}, that are not exactly about quantum sensing but are closely related in that they concern how to most efficiently extract information about a quantum system.  For example, quantum hypothesis testing is used to determine optimal schemes for the discrimination of quantum states or quantum channels~\cite{helstrom1969quantum}. If these states or channels are taken to encode information about a physical signal through a sensing interaction, sensing tasks can be modeled as discrimination problems to which the framework of quantum hypothesis testing can be applied~\cite{banchi_quantum-enhanced_2020, gardner_quantum_2026}. Quantum learning protocols where the learning objective could similarly be modeled to encode a sensing signal include schemes for learning properties of quantum states~\cite{Huang_2022, dimitroff_learning_2026}, learning properties of Hamiltonians~\cite{huang2023learning, ott_hamiltonian_2024, chen_learning_2025}, deciding whether a function that can be accessed using a quantum oracle satisfies or does not satisfy a specific property (referred to as quantum property testing~\cite{montanaro_survey_2016}), among others.

Developing QCS protocols can be viewed as an exercise in discovering how to use quantum computation to enhance the performance of quantum sensors for specific tasks. From this perspective, it is natural to consider which of the many quantum algorithms and quantum algorithmic primitives that have been developed in quantum computing can be adapted to give advantages for sensing tasks---just as we have seen with a few examples already based on Grover's algorithm, quantum signal processing, and quantum neural networks. Could other oracle-based quantum algorithms~\cite{deutsch1992rapid,bernstein1993quantum,simon1997power}, or quantum-linear-algebra algorithms such as the HHL algorithm~\cite{harrow_quantum_2009}, be adapted to provide an advantage in quantum computational sensing? Quantum-machine-learning algorithms such as quantum reservoir computing~\cite{fujii2021quantum, mujal2021opportunities} could potentially be applied to a wide variety of tasks. Ref.~\cite{krisnanda2022phase} takes a step in this direction, although its motivation is conventional phase sensing: it describes a protocol where a quantum reservoir processes a quantum state that encodes a phase and produces functions of the sensed phase. A linear layer executed in classical post-processing is then used to produce an estimate of the sensed phase. However, the same features could plausibly be used to obtain a QCSA by constructing other linear layers trained for specific tasks. Quantum reservoir computing has also been used experimentally to classify radio-frequency signals impinging on a quantum system \cite{senanian2024microwave}; it appears likely that quantum reservoir computing could similarly be applied with quantum sensors of radio-frequency displacements~\cite{backes2021quantum} to achieve a QCSA. 

\section{Summary and Outlook}
\label{sec:outlook}

In this Perspective, we have described how quantum sensors can be combined with quantum computing to form quantum computational sensors, and how these can be used to achieve a new kind of quantum advantage: a quantum computational-sensing advantage (QCSA). The opportunity for an advantage comes about when designing a system whose goal is not to report raw values of sensed physical quantities, but to perform some more sophisticated task that is a function of underlying sensed quantities---which themselves don't need to be reported. Two general and complementary ways of intuitively understanding how a QCSA can arise are as follows. First is the \emph{information perspective}: measuring a qubit causes its collapse, and as a result it is only possible to extract a single classical bit of information from each qubit in a sensing system\footnote{This notion of quantum measurements resulting in only a relatively small amount of classical information being obtained extends to the case of qudits, and of qumodes (bosonic modes).}, in a single shot. One can gain an advantage for a specific task by quantum computing within the sensor prior to measurement so that the limited classical information revealed by measurement is as relevant to the task one is trying to perform as possible. Second is the \emph{noise perspective}: measurement of quantum systems in general results in highly stochastic---i.e., noisy---outcomes. It may be possible to quantum compute before measurement so that the measurement result is less noisy and more directly relevant to the task one is ultimately trying to perform, i.e., the function $\Ft$ one is trying to approximate. Having the quantum system output estimates of the function $\Ft(\bm{u})$ directly by computing it in the quantum system before measurement can be superior to reading out noisy estimates of the raw sensed values $\bm{u}$ and computing the function classically in postprocessing because of how noise propagates \cite{tellinghuisen2001statistical}: for many nonlinear functions, the signal-to-noise ratio\footnote{Many of the QCS tasks we have reviewed are discrimination or classification tasks, for which metrics such as the true positive rate and false positive rate, or classification accuracy, are ultimately more relevant than the signal-to-noise ratio, but higher signal-to-noise ratio is desirable because it generally enables discrimination or classification to be performed with higher accuracy.} will be degraded by the postprocessing. Even for some linear functions of $\bm{u}$, there can be a large benefit to having the quantum system output $\Ft(\bm{u})$ directly. The information perspective and the noise perspective can each give helpful ways to reason about how a quantum computational sensor can achieve an advantage for a particular task (function), but they are also intimately related through the connection between mutual information and signal-to-noise ratio \cite{cover2006elements}.

\subsection{Connection with quantum machine learning}

QCS encompasses tasks and approaches that have no connection with machine learning---as several of the examples we have covered (e.g., in Fig.~\ref{fig:refs} and Fig.~\ref{fig:schemes}), such as codeword discrimination, threshold detection, trajectory sensing, and Grover-based signal detection, show. However, we have also cited and discussed several QCS examples based on machine-learning approaches, and believe that there is a natural opportunity to further explore the use of quantum machine learning (QML) in quantum computational sensors. Attempts to apply QML to classical data have suffered from numerous challenges \cite{cerezo2022challenges}, including how to---even in principle---achieve a practical quantum advantage when the input classical data is small, and how in practice to load large classical data efficiently \cite{aaronson2025future}. QCS can be thought of as quantum computing on classical data that happens to be unknown (the sensed parameters) or as quantum computing on quantum data that the quantum sensors produce. In the first framing, the data is classical and small\footnote{If the quantum computational sensor has $M$ constituent qubits/qumodes/etc., it will typically sense at most $M$ classical parameters---in other words, a number linear rather than exponential in the system size---and in most settings we can imagine for quantum sensors, $M$ will be small by machine-learning standards (less than $10^6$, and typically much less in the near and intermediate term).}. Yet even so, we have seen examples where quantum-machine-learning methods such as quantum neural networks can enable a \textit{computational-sensing} advantage---where a \textit{purely computational} advantage (if the classical parameters were known exactly a priori) would be implausible. QCS avoids the large-classical-data regime of QML, sidestepping concerns about the practicality of quantum random access memory (QRAM) for loading classical data into quantum computers, which have been a source of much consternation in the search for purely computational advantages with QML \cite{aaronson2025future}.

In the next subsection we discuss open questions about the advantage one might achieve, both with QML and non-QML approaches.

\subsection{How much advantage is possible, for what tasks, and using what protocols?}

We have discussed a variety of examples of quantum computational sensors that can achieve a QCSA for specific tasks (including those in Figs.~\ref{fig:refs} and \ref{fig:schemes}) and have seen that the advantages can be by constant factors or can scale as a function of the sensing resources (such as the number of photons used) or the signal strength. Some of the advantages that have been discovered so far are already large enough that they seem like they could become practically important. For example, Ref.~\cite{chin2024quantumentanglementenablessingleshot} (Fig.~\ref{fig:schemes}d) reports, for the trajectory-sensing task it considers, in some range of signal strengths, an error probability of exactly 0\% with a quantum computational sensor, whereas using conventional quantum sensing, the error probability would be close to 20\%. Finding for any given task how much advantage is possible, and with what quantum-computational-sensor design, is a major open research direction. Advantages that are exponential in the number of qubits or qumodes involved are plausible, in part due to the following connection between QCS and recent work on quantum algorithmic measurement \cite{Aharonov_2022} and learning from experiments \cite{Huang_2022}, where exponential advantages have been proven for specific discrimination and estimation tasks: both Refs.~\cite{Aharonov_2022} and \cite{Huang_2022} consider the setting of attempting to learn something about an unknown quantum state, and this is also a common setting in quantum machine learning of quantum data \cite{cerezo2022challenges}. A subset of possible QCS protocols can be framed as: a quantum sensor (or system of quantum sensors) first produces a quantum state that encodes the sensed signal $\bm{u}$, and then quantum processing and measurement is used to extract the relevant information from the state to obtain an estimate of $\Ft(\bm{u})$---so the techniques of Refs.~\cite{Aharonov_2022} and \cite{Huang_2022} also apply to quantum sensing, raising the open questions of which practical sensing tasks they may be adapted for and whether exponential QCSA is achievable for them.

\definecolor{color_m}{RGB}{55, 171, 200} 
\definecolor{color_e}{RGB}{68, 120, 33} 
\definecolor{color_c}{RGB}{200, 113, 55} 

\begin{table}[t]
    \centering
    \begin{adjustbox}{width=1.0\textwidth,center=\textwidth} 
    \begin{tabular}{|c |c |r |} 
    \hline
    \rule{0pt}{4ex}\large{\textbf{Quantum Algorithm / Algorithmic Primitive}} & \large{\textbf{Examples}} & \large{\textbf{Scheme}} \\ [0.5ex] 
    \hline

    \multicolumn{3}{|c|}{\rule{0pt}{4ex}\large{\textbf{Examples of quantum computational sensing (QCS)}}} \\ [0.5ex] \hline

    \rule{0pt}{3ex}
    Belief propagation with quantum messages&
    
    \begin{tabular}{@{}c@{}}
    Joint detection (detecting codewords)~\cite{delaney2022demonstration, rengaswamy_belief_2021, renes_belief_2017}
    \end{tabular}&

    \begin{tabular}{@{}c@{}}
    Analytic \raisebox{-0.3ex}{\tikz \fill[color_m] (0,-1) circle [radius=0.15cm];}
    \end{tabular} \\ [0.5ex]
    \hline

    \rule{0pt}{3ex}
    Walsh--Hadamard transform&
    
    \begin{tabular}{@{}c@{}}
    Joint detection (detecting codewords)~\cite{guha2011quantum}
    \end{tabular}&

    \begin{tabular}{@{}c@{}}
    Analytic \raisebox{-0.3ex}{\tikz \fill[color_m] (0,-1) circle [radius=0.15cm];}
    \end{tabular} \\ [0.5ex]
    \hline

    \rule{0pt}{3ex}
    Quantum Fourier transform&
    
    \begin{tabular}{@{}c@{}}
    Quantum-assisted telescope arrays (measuring stellar intensity distributions)~\cite{gottesman_longer-baseline_2012, khabiboulline_optical_2019, khabiboulline_quantum-assisted_2019}
    \end{tabular}&

    \begin{tabular}{@{}c@{}}
    Analytic \raisebox{-0.3ex}{\tikz \fill[color_m] (0,-1) circle [radius=0.15cm];}
    \end{tabular} \\ [0.5ex]
    \hline

    \rule{0pt}{3ex}
    Measurement basis optimization&
    
    \begin{tabular}{@{}c@{}}
    Quantum enhanced receivers (discriminating coherent states)~\cite{tsujino2011quantum, han2020helstrom, sasaki1996optimum, warke2024photonic} \\
    Quantum enhanced receivers (discriminating multiple copies of coherent states)~\cite{burenkov2021practical, jabir2020experimental, becerra2013experimental, becerra2015photon, dolinar1976class} \\
    Superresolution imaging~\cite{tsang2016quantum, jha2025multi, grace2020approaching, lee2022quantum, wang_advancing_2025} \\
    Quantum illumination receiver (detecting targets)~\cite{jo2021quantum, cox2024quantum} \\
    Nonlinear bosonic amplifiers (estimating functions of displacement)~\cite{khan_quantum_2025} \\
    Directional detection of dark matter axions~\cite{fukuda2025directional} \\
    Joint detection (detecting codewords)~\cite{chen2012optical, guha_structured_2011, crossman2024quantum, sarovar_quantum_2023} \\
    Quantum-enhanced optical imaging~\cite{mokeev_enhancing_2025} \\ 
    Quantum superresolution and noise spectroscopy (detection of weak incoherent signals)~\cite{de_almeida_superresolving_2025, gardner_quantum_2026} \\ 
    Photon-counting optical interferometers (detection of stochastic signals)~\cite{mcculler_single-photon_2022}
    \end{tabular}&

    \begin{tabular}{@{}c@{}}
    Analytic \raisebox{-0.3ex}{\tikz \fill[color_m] (0,-1) circle [radius=0.15cm];}
    \end{tabular} \\ [0.5ex]
    \hline

    \rule{0pt}{3ex}
    Parameterized measurement basis optimization&
    
    \begin{tabular}{@{}c@{}}
    Quantum enhanced receivers (discriminating multiple copies of coherent states)~\cite{cui2022quantum}
    \end{tabular}&

    \begin{tabular}{@{}c@{}}
    Learned \raisebox{-0.3ex}{\tikz \fill[color_m] (0,-1) circle [radius=0.15cm];}
    \end{tabular} \\ [0.5ex]
    \hline

    \rule{0pt}{3ex}
    Probe state and measurement basis optimization&
    
    \begin{tabular}{@{}c@{}}
    Quantum sensor networks (estimating weighted sum of parameters)~\cite{proctor2018multiparameter, guo2020distributed, rubio2020quantum, bringewatt2024optimal, ge2018distributed, eldredge2018optimal, bringewatt2021protocols, xia_entanglement-enhanced_2023, malia_distributed_2022, van2024utilizing} \\
    Quantum sensor networks (estimating functions of parameters)~\cite{qian_heisenberg-scaling_2019, qian2021optimal, fromonteil_non-local_2025} \\
    Quantum sensor networks (discrete outcome detection)~\cite{hillery_discrete_2023, ali_discrete-outcome_2024} \\
    Bosonic QSP interferometry (classifying 1D displacement)~\cite{sinanan-singh_single-shot_2024} \\
    Pattern recognition~\cite{banchi_quantum-enhanced_2020, ortolano2023quantum} \\
    Quantum trajectory sensing~\cite{chin2024quantumentanglementenablessingleshot} \\
    Quickest change detection of transmission loss~\cite{guha2025quantum} \\
    Quantum nonlinear spectroscopy (measuring higher-order correlations of phase)~\cite{meinel2022quantum} \\
    Thresholded quantum sensing (threshold detection of second-order moments of parameters)~\cite{alderete2025nonlinear} \\
    Quantum signal learning (estimating many  properties of a classical signal)~\cite{cotler_quantum_2026, kannan2026exponential}\\
    Broadband signal detection~\cite{polloreno_simple_2026}    \\
    Stochastic sensing (estimating properties of probability distributions of signals)~\cite{wang2024exponential,prabhu2025exponential}    \\
    Quantum error correction-enhanced logical sensing (detecting weak pulse sequences)~\cite{ott_rare-event_2026}
    \end{tabular}&

    \begin{tabular}{@{}c@{}}
    Analytic \raisebox{-0.3ex}{\tikz \shade[left color=color_e, right color=color_m, shading angle=45] (0,-1) circle [radius=0.15cm];}
    \end{tabular} \\ [0.5ex]
    \hline

    \rule{0pt}{3ex}
    Parameterized probe state and measurement basis optimization&
    
    \begin{tabular}{@{}c@{}}
    Quantum sensor networks (locating transmitter)~\cite{zhan2023quantum} \\
    Supervised learning assisted by an entangled sensor network (SLAEN) \\ (estimating weighted sum of displacements)~\cite{zhuang_physical-layer_2019, xia_quantum-enhanced_2021} \\ 
    Nonlinear SLAEN (classifying 2D displacement)~\cite{liao_quantum-enhanced_2024} \\
    Bosonic QSP interferometry (classifying 2D displacement)~\cite{khan_quantum_2025, prabhu_quantum_2026} \\
    Global function sensing (estimating nonlinear functions)~\cite{ilias_end--end_2025} \\
    Variational multimode photon counting (learning statistical properties of weak signals)~\cite{feng_physical-layer_2025} \\
    Quantum-correlation-enhanced imaging (low-light image classification)~\cite{sohoni_ultra-low-light_2026, shi_resource-efficient_2026}
    \end{tabular}&

    \begin{tabular}{@{}c@{}}
    Learned \raisebox{-0.3ex}{\tikz \shade[left color=color_e, right color=color_m, shading angle=45] (0,-1) circle [radius=0.15cm];}
    \end{tabular} \\ [0.5ex]
    \hline

    \rule{0pt}{3ex}
    Grover's algorithm&
    
    \begin{tabular}{@{}c@{}}
    Detecting AC signal with unknown frequency and amplitude~\cite{allen2025quantumcomputingenhancedsensing}
    \end{tabular}&

    \begin{tabular}{@{}c@{}}
    Analytic \raisebox{-0.3ex}{\tikz \fill[color_c] (0,-1) circle [radius=0.15cm];}
    \end{tabular} \\ [0.5ex]
    \hline

    \rule{0pt}{3ex}
    Quantum signal processing (QSP)&
    
    \begin{tabular}{@{}c@{}}
    Qubit QSP (classifying phases)~\cite{debry_experimental_2023, khan_quantum_2025, sen_qsp_2026}
    \end{tabular}&

    \begin{tabular}{@{}c@{}}
    Learned \raisebox{-0.3ex}{\tikz \fill[color_c] (0,-1) circle [radius=0.15cm];}
    \end{tabular} \\ [0.5ex]
    \hline

    \rule{0pt}{3ex}
    Quantum neural networks (QNN)&
    
    \begin{tabular}{@{}c@{}}
    Qubit QNN (classifying spatio-temporal phase)~\cite{khan_quantum_2025}
    \end{tabular}&

    \begin{tabular}{@{}c@{}}
    Learned \raisebox{-0.3ex}{\tikz \fill[color_c] (0,-1) circle [radius=0.15cm];}
    \end{tabular} \\ [0.5ex]
    \hline

    \multicolumn{3}{|c|}{\rule{0pt}{4ex}\large{\textbf{Examples of protocols for learning about quantum systems, which could potentially be extended to QCS}}} \\ [0.5ex]
    \hline

    \rule{0pt}{3ex}
    Classical shadows&
    
    \begin{tabular}{@{}c@{}}
    Classical approximation of quantum states from few samples~\cite{huang2020predicting, aaronson2018shadow, bringewatt_classical_2026}
    \end{tabular}&

    \begin{tabular}{@{}c@{}}
    Analytic \raisebox{-0.3ex}{\tikz \fill[color_m] (0,-1) circle [radius=0.15cm];}
    \end{tabular} \\ [0.5ex]
    \hline

    \rule{0pt}{3ex}
    Non-local measurement across copies&
    
    \begin{tabular}{@{}c@{}}
    Quantum-enhanced learning (classifying quantum states)~\cite{Huang_2022, conlon_discriminating_2023, noller_infinite_2025}
    \end{tabular}&

    \begin{tabular}{@{}c@{}}
    Analytic \raisebox{-0.3ex}{\tikz \fill[color_m] (0,-1) circle [radius=0.15cm];}
    \end{tabular} \\ [0.5ex]
    \hline

    \rule{0pt}{3ex}
    Quantum convolutional neural networks&
    
    \begin{tabular}{@{}c@{}}
    Classifying topological phases of quantum states~\cite{cong2019quantum}
    \end{tabular}&

    \begin{tabular}{@{}c@{}}
    Learned \raisebox{-0.3ex}{\tikz \fill[color_m] (0,-1) circle [radius=0.15cm];}
    \end{tabular} \\ [0.5ex]
    \hline

    \rule{0pt}{3ex}
    Entangling unitaries&
    
    \begin{tabular}{@{}c@{}}
    Learning Pauli channels~\cite{Chen_2022} \\
    Learning bosonic displacement channels~\cite{oh2024entanglement}
    \end{tabular}&

    \begin{tabular}{@{}c@{}}
    Analytic \raisebox{-0.3ex}{\tikz \shade[left color=color_e, right color=color_m, shading angle=45] (0,-1) circle [radius=0.15cm];}
    \end{tabular} \\ [0.5ex]
    \hline

    \rule{0pt}{3ex}
    SWAP / Hadamard Test&
    
    \begin{tabular}{@{}c@{}}
    Quantum algorithmic measurement (classifying ensemble of unitaries)~\cite{Aharonov_2022}
    \end{tabular}&

    \begin{tabular}{@{}c@{}}
    Analytic \raisebox{-0.3ex}{\tikz \fill[color_c] (0,-1) circle [radius=0.15cm];}
    \end{tabular} \\ [0.5ex]
    \hline

    \rule{0pt}{3ex}
    Amplitude amplification&
    
    \begin{tabular}{@{}c@{}}
    Quantum enhanced estimation of observables~\cite{wada2024quantum}
    \end{tabular}&

    \begin{tabular}{@{}c@{}}
    Analytic \raisebox{-0.3ex}{\tikz \fill[color_c] (0,-1) circle [radius=0.15cm];}
    \end{tabular} \\ [0.5ex]
    \hline

    \rule{0pt}{3ex}
    Hamiltonian and quantum state learning&
    
    \begin{tabular}{@{}c@{}}
    Learning many-body Hamiltonians and states~\cite{huang2023learning, ott_hamiltonian_2024,chen_learning_2025, dimitroff_learning_2026}
    \end{tabular}&

    \begin{tabular}{@{}c@{}}
    Analytic \raisebox{-0.3ex}{\tikz \fill[color_c] (0,-1) circle [radius=0.15cm];}
    \end{tabular} \\ [0.5ex]
    \hline

    \rule{0pt}{3ex}
    Quantum hypothesis testing&
    
    \begin{tabular}{@{}c@{}}
    Discriminating quantum states or quantum channels~\cite{helstrom1969quantum}
    \end{tabular}&

    \begin{tabular}{@{}c@{}}
    Analytic \raisebox{-0.3ex}{\tikz \fill[color_e] (0,-1) circle [radius=0.15cm];}
    \end{tabular} \\ [0.5ex]
    \hline

    \rule{0pt}{3ex}
    Quantum property testing&
    
    \begin{tabular}{@{}c@{}}
    Deciding if a function satisfies or does not satisfy a given property~\cite{montanaro_survey_2016}
    \end{tabular}&

    \begin{tabular}{@{}c@{}}
    Analytic \raisebox{-0.3ex}{\tikz \fill[color_c] (0,-1) circle [radius=0.15cm];}
    \end{tabular} \\ [0.5ex]
    \hline

    \end{tabular}
    \end{adjustbox}
    
    \vspace{1em}
    
    \begin{adjustbox}{width=0.5\textwidth}
    \begin{tabular}{|c|c|}
        \hline
        \multicolumn{2}{|c|}{\rule{0pt}{4ex}\large{\textbf{Legend}}}
        \\ [0.5ex] \hline

        \rule{0pt}{3ex}
        Single sensing operation per measurement shot with measurement basis optimization & \raisebox{-0.7ex}{\tikz \fill[color_m] (0,-1) circle [radius=0.15cm];} \\ [0.5ex] \hline

        \rule{0pt}{3ex}
        Single sensing operation per measurement with probe state and measurement basis optimization & \raisebox{-0.7ex}{\tikz \shade[left color=color_e, right color=color_m, shading angle=45] (0,-1) circle [radius=0.15cm];} \\ [0.5ex] \hline

        \rule{0pt}{3ex}
        Multiple sensing operations per measurement shot & \raisebox{-0.7ex}{\tikz \fill[color_c] (0,-1) circle [radius=0.15cm];} \\ [0.5ex] \hline
        
    \end{tabular}
    \end{adjustbox}

    \caption{\textbf{Quantum algorithms and quantum algorithmic primitives that have already been used in quantum computational sensing, or in learning properties of quantum states.} The examples are categorized by whether the target function is analytic or learned, and the structure the protocol has, based on the definitions in Fig.~\ref{fig:taxo} (see legend).}
    \label{table:examples_qcsa}
\end{table}

One situation where this can arise is if the sensed parameters $\bm{u}$ are stochastic and the goal is to estimate a function $\Ft(\bm{u})$ that determines a property of the underlying probability distribution~(see Refs.~\cite{wang2024exponential,prabhu2025exponential, kannan2026exponential}). For tasks requiring for example, classifying which distribution the parameters were drawn from, or estimating quantities that govern their probability distribution, quantum computational sensing protocols using entanglement can be designed to obtain an exponential advantage over unentangled quantum sensors that estimate the stochastic parameters individually. The QCSA exceeds the usual polynomial advantage of Heisenberg versus Standard-Quantum-Limit scaling~\cite{Degen_2017} when sensing deterministic parameters. An interesting direction for future work is to explore which probability distributions and task functions---especially combinations of the two that might have practical applications---allow an exponential, or other very large, QCSA.

We listed in Table~\ref{table:examples_qcsa} a variety of examples of quantum algorithmic techniques, including approaches from quantum machine learning, that have been or could be adapted to QCS. Since some of these quantum algorithms provide provable computational speedups for their respective tasks, it is plausible that they might enable similar (e.g., polynomial or exponential) QCSA. For instance, we highlighted in Sec.~\ref{sec:Grover} the use of Grover's algorithm as a computational routine for AC-field detection, framed as a search problem~\cite{allen2025quantumcomputingenhancedsensing}. In a purely computational setting, Grover's algorithm gives a quadratic speedup~\cite{grover_fast_1996}. In the QCS setting, the Grover-based approach gave a QCSA that was polynomial but not quadratic. This example underscores the ability to use quantum algorithms that have a computational scaling advantage to obtain a QCSA that also scales (with system size, sensing resources, task size, etc.)---but that the quantitative scaling relation might not be identical to what one obtains in the purely computational setting. Ultimately, it is an open challenge to discover which quantum algorithms and algorithmic primitives are useful for quantum computational sensing, for which tasks they are useful, how they can best be incorporated into full quantum-computational-sensing protocols, and how much advantage the protocols can achieve.

\subsection{Experimental demonstrations}


We will now discuss initial proof-of-principle demonstrations of quantum computational-sensing advantage using QCS protocols realized in experiments.

\subsubsection{Linear function estimation}

QCS protocols for the task of estimating linear functions of distinct sensed signals have already been demonstrated using quantum sensor networks~(Fig.~\ref{fig:schemes}a). Ref.~\cite{guo2020distributed} reports on a demonstration of a bosonic quantum sensor network used to compute a specific function: the average of the phase shifts experienced by the individual modes, $\Ft = \frac{1}{M}\sum_i \theta_i$. Starting with a single-mode squeezed state generated using an optical parametric amplifier, the QCS protocol requires preparing an entangled probe state of $M=4$ spatial modes of a propagating field using balanced beam splitters. The modes then experience local phase shifts that constitute the sensing operation. Homodyne measurements of the individual modes are then performed and the results used to estimate the average phase shift. The conventional QS strategy instead prepares single-mode squeezed probe states to estimate the phase experienced by each spatial mode, following by classical postprocessing to estimate the average. For the same probe state photon number, the QCS protocol of Ref.~\cite{guo2020distributed} exhibits a sensitivity gain (increase in square root of signal-to-noise ratio) of $0.76$~dB over the QS strategy.

Quantum sensor networks for estimating averages of sensed signals have also been demonstrated in discrete-variable platforms. The QCS protocol in Ref.~\cite{liu_distributed_2021} uses polarization-entangled states of six spatial modes of an optical field followed by six-photon coincidence measurements to directly estimate the average phase shift experienced by a subset of the modes. Compared to the QS strategy that uses separable modes to first estimate the individual phase shifts before computing the desired average classically, this QCS protocol achieves a sensitivity gain of $4.7$~dB. Ref.~\cite{malia_distributed_2022} describes a QCS protocol that is used to estimate the average phase shift experienced across spatial modes of atomic ensembles. Prior to sensing, the protocol generates an entangled network of up to four spin-momentum modes of atomic ensembles with collective spin-squeezing, and then measures the collective spin to directly extract the average phase shift. This QCS strategy achieves a sensitivity gain of $4.5$~dB over the QS strategy that uses separable spin-squeezed states of the modes to estimate individual phases.

Finally, quantum sensor networks have also been demonstrated for the estimation of linear functions beyond the computation of an average value. In Ref.~\cite{xia_demonstration_2020}, the authors demonstrate the estimation of the phase gradient across a network of three spatial optical modes, which can be achieved by computing a three-point finite-differences function $\Ft = -\frac{3}{2}\theta_1 + 2\theta_2 - \frac{1}{2}\theta_3$. By preparing an appropriate entangled probe state, the experimental QCS protocol provides an estimated $0.25$~dB sensitivity gain over the conventional QS protocol that only uses a local squeezed probe state to estimate each individual phase.

\subsubsection{Classification}

QCS protocols have also been experimentally demonstrated for classification tasks using the SLAEN approach (Fig.~\ref{fig:schemes}b) in Ref.~\cite{xia_quantum-enhanced_2021}. The authors use a network of three spatial bosonic modes to perform binary classification of data encoded in the amplitudes and phases of received radiofrequency (RF) signals. The conventional QS strategy would prepare single mode squeezed probe states in each mode to estimate each received signal, followed by a classical postprocessor for classification. Using an optimized variational quantum circuit to engineer an entangled probe state prior to sensing, the SLAEN approach can perform QCS to more accurately extract features relevant for classification. The resulting protocol is estimated to enable a 13\% reduction in classification error compared to the QS strategy under the constraint of a fixed probe state photon number.

The Quantum Signal Processing (QSP)~\cite{martyn2021grand} algorithm has been used in QCS protocols for classification in Ref.~\cite{debry_experimental_2023}. The authors consider the task of discriminating quantum channels analogous to phase and amplitude encodings of signals in classical radio communication using a single trapped ion sensor. Using multiple accesses to the channels (akin to multiple sensing periods) interleaved with QSP operations, the QCS protocol achieves single-shot discrimination with $99\%$ accuracy, a $17\%$ reduction in classification error over the best-performing QS strategy using only a single channel access per measurement.

The QSP algorithm has also been used in QCS protocols for classification of real magnetic fields in Ref.~\cite{sen_qsp_2026}. The authors consider magnetic field sensing using a single transmon qubit, by estimation of the phase rotation imparted by the incident field on the qubit state. Several conventional approaches exist for phase estimation using qubits, from Ramsey interferometry to quantum phase estimation and dynamical-decoupling schemes. The authors demonstrate the alternative QCS approach that implements variationally-optimized single-qubit gates inspired by QSP to coherently process the sensed field and directly extract class label information from qubit measurement. For a fixed total protocol time, the QCS approach outperforms conventional QS baselines for both static and time-varying magnetic field sensing by 15 and 20 percentage points respectively.

Hybrid qubit-cavity sensors (Fig.~\ref{fig:schemes}c) have been demonstrated for QCS of displacements~\cite{prabhu_quantum_2026}. In this work, the cavity mode is used as a bosonic sensor, receiving an RF signal that imparts a displacement on the mode; only a single sensing operation is allowed. The conventional QS strategy uses a phase-preserving amplifier to perform heterodyne measurement of both quadratures of the received displacement, and a classical postprocessor to determine the class label. The QCS protocol instead uses a qubit coupled to the bosonic sensor to optimize the joint state before and after the sensing step, such that a measurement of the qubit directly outputs the class label. Given a single measurement shot, the QCS protocol achieves an advantage of up to 15\% points in classification accuracy over the conventional QS strategy.

A similar hybrid qubit-cavity sensor has been used to demonstrate an exponential sample advantage in sensing of stochastic displacements over conventional Gaussian sensing protocols~\cite{kannan2026exponential}. In this work, the goal is to learn features of the probability distribution governing the displacements imparted on the bosonic sensor, such as Fourier components and temporal correlations. Coupling a computational qubit to the bosonic sensor allows performing a QCS protocol that directly extracts a specific Fourier component or temporal correlation function of interest via measurement of the qubit alone. This protocol is demonstrated to learn features with exponentially fewer samples than Gaussian protocols using equivalent energy, achieving more than $10^7$-fold reduction in samples in experiment.

Finally, QCS protocols for classification tasks have also been demonstrated using quantum-enhanced receivers. In Ref.~\cite{cui2022quantum}, the authors demonstrate an optical quantum receiver enhanced by adaptive learning (QREAL) to discriminate coherent states encoding received signals. After receiving the coherent state in a fiber-based optical mode, the QREAL architecture applies multiple trainable processing operations via displacements of the optical field, interleaved with partial measurements via photon number resolving detection. The prior measurement history is used to adaptively adjust subsequent processing operations. Across two coherent-state encoding schemes, this adaptive receiver provides a $40\%$ reduction in error rate compared to the conventional QS strategy using lossless quadrature measurement, and a $14\%$ reduction in comparison to optimal Dolinar receivers. 

Quantum-enhanced receivers for discriminating coherent states can also be implemented in qubit systems, provided the received states can first be efficiently transduced to qubit states. Minus this transduction step, the qubit-enabled computing backend of such receivers has been demonstrated in a trapped-ion quantum processor in Ref.~\cite{delaney2022demonstration}, using a three-qubit quantum circuit. The authors predict that the demonstrated processor, if combined with a sensing and transduction step to implement a true receiver, would be able to achieve a lower error rate than the conventional QS strategies of both homodyne measurement and optimal Helstrom measurements for coherent state discrimination at a fixed received photon number. A variationally-optimized computing backend has also been demonstrated using a superconducting quantum processor in Ref.~\cite{crossman2024quantum}. When combined with a sensing and transduction module, the resulting performance is again predicted to be able to outperform Helstrom measurements for discriminating codewords of three coherent states at a fixed photon number.  \\

These demonstrated QCS protocols compute only linear functions of sensed signals, and are in this sense on the simpler end of QCS schemes. Several obstacles remain for the demonstration of advantages using protocols to compute more complicated functions.

\subsection{Experimental realization: challenges to proof-of-principle demonstrations and to achieving practical advantage}

We now explain some of the challenges that need to be addressed to experimentally realize some of the more complicated quantum computational-sensing protocols, as well as to construct quantum computational sensors that deliver practically-relevant advantages.

\subsubsection{Proof-of-principle demonstrations of advantage}

A central challenge is how to handle hardware imperfections, especially decoherence. Any physical realization of quantum hardware suffers from decoherence and loss, so there are strong practical constraints on how much fidelity one can expect a system's quantum state to maintain, especially when running longer protocols (deeper circuits). Decoherence and loss can be deleterious both by degrading the effectiveness of the metrological probe states that are used for sensing within the QCS protocol, and by causing errors in the quantum computation that may reduce the accuracy of estimating the target function $\Ft$. Conventional quantum sensing protocols already suffer from the former: entangled states are less robust against noise than unentangled states, and their enhanced sensitivity to signals can---as is the case for GHZ states---even be exactly negated by their increased decoherence rate \cite{Degen_2017}. In QCS, the challenge of dealing with decoherence to entangled states can be even more severe: several of the protocols we have described (e.g., in Fig.~\ref{fig:schemes}) involve more than one period of sensing as well as additional circuit depth for computation, so there is a longer time for which the system needs to maintain its sensing benefit from entanglement in the presence of decoherence, versus conventional protocols, which often involve only one period of sensing and low-depth circuits on either side of it. In addition to considering how to keep the sensing part of a quantum computational sensor effective in the presence of decoherence, one also has to make sure the advantage survives given that the computation part will also suffer from decoherence---and this latter challenge is one that doesn't apply to most conventional quantum sensors\footnote{Some conventional quantum sensors---which are conventional in that they still aim to return raw sensed parameters $\bm{u}$ rather than a function $\Ft(\bm{u})$ of them---do rely on quantum computation, such as performing a Quantum Fourier Transform \cite{Degen_2017}, and also need to ensure that the computation part of their quantum protocol survives the decoherence.}. A major challenge in experimentally demonstrating QCSA for protocols that have deep circuits is ensuring that they can still, in the presence of decoherence, deliver their advantage over a conventional quantum sensor that may use a much shallower circuit.

How can one approach this challenge? Besides hoping for continued improvements in the physical error rates of quantum-hardware platforms, we anticipate that it will be useful to explore adopting four other broad strategies from conventional quantum sensing: using quantum error correction \cite{Degen_2017}; for protocols involving the readout of expectation values, using quantum error mitigation \cite{yamamoto2022error} and inference-based sensing \cite{huerta2022inference}; for protocols involving entanglement, exploring the use of entangled states that are more robust to noise \cite{Degen_2017}; and for QCS protocols involving tunable parameters (such as ones based on quantum signal processing and quantum neural networks), using the approach from variational metrology to optimize the tunable parameters in the presence of the hardware imperfections \cite{marciniak_optimal_2022}. Just as with noisy, intermediate-scale quantum (NISQ) \cite{preskill2018quantum} computing, it is also natural to work on reducing the circuit depth and gate count of protocols as much as possible.

An important next step for many of the nascent QCS protocols we have covered is to analyze their potential for delivering an advantage in the presence of realistic hardware imperfections, and to introduce protocol variants whose advantage is more robust to hardware imperfections.

\subsubsection{Practical advantage}

A natural first step for experimental realization of each QCS protocol is to show a proof-of-concept that the protocol can provide an advantage when compared with a conventional quantum sensing protocol run on exactly the same physical hardware. QCSA is defined in Box~2 in this way, and this is similar to how advantages in conventional quantum sensing are often defined: typically advantages are demonstrated using different protocols run on the same hardware, such as a protocol using entanglement versus not \cite{Degen_2017}. However, ultimately what we would like is for an implementation of a quantum computational sensor to deliver an advantage, for a task of practical interest, not merely over a simpler protocol run on the same hardware, but over all other solutions to performing the task that the quantum computational sensor performs, including the best other quantum and classical sensors, under some set of constraints on size, weight, power, robustness to environmental conditions (such as operating temperature and vibrations), uptime, maintainability, toxicity, cost (\$) etc.---which would be not just a proof-of-concept advantage, but a \textit{practical advantage}.

Engineering \emph{conventional} quantum sensors to achieve performance better than what is possible with state-of-the-art classical sensors in practical scenarios is a major open challenge but nevertheless has already been achieved for a small number of applications and sensing platforms \cite{bongs2023quantum,gschwendtner2024quantum,muradoglu2025quantum}. The constraints of practical applications often preclude the use of some of the most prominent quantum hardware platforms---for example, there are very few sensing applications where it is practical for the sensor to need to be inside a dilution refrigerator.\footnote{For some applications, needing a dilution refrigerator can be acceptable though---for example, in dark-matter searches \cite{backes2021quantum,agrawal2024stimulated,braggio2025quantum}.} It is also important to choose a quantum hardware platform that has state-of-the-art sensitivity, which will again preclude some prominent platforms that have other positive attributes but are not especially sensitive.

Achieving practical advantage for quantum \emph{computational} sensors will require similar effort in selecting and further engineering hardware platforms to satisfy the practical constraints while also delivering state-of-the-art sensitivity. However, there is the added consideration of needing the quantum system to be capable of performing the quantum computation called for in the QCS protocol that one hopes to execute on it---which in general makes practical advantage with a quantum computational sensor even more difficult to achieve than practical advantage with a conventional quantum sensor. One challenge here is broadly the same as in the discussion of challenges for experimental proof-of-principle demonstrations of QCSA: how to deal with decoherence. A quantum sensor that has excellent sensitivity (and so might seem like a good starting point for a quantum computational sensor) won't necessarily also have sufficiently low error rates to be able to successfully run a given QCS protocol. Proof-of-principle demonstrations of QCSA can be done with hardware platforms that don't have state-of-the-art sensitivity but do have very low error rates, since the aim is to show an advantage from computational sensing relative to conventional sensing given access to the same resources. But to achieve practical advantage, it is important to choose hardware that has excellent sensitivity so that the quantum computational sensor can deliver an advantage over \emph{any} other sensing system. A second challenge that can be sidestepped in proof-of-principle demonstrations but that must be addressed to realize practical advantage is that hardware platforms that have excellent sensitivity might not have a sufficient number of qubits (or qumodes), or sufficient controllability, to be able to execute the QCS protocol. For example, nitrogen-vacancy (NV) centers in diamond can be used as state-of-the-art magnetic-field sensors \cite{Degen_2017}, but each NV center typically gives access to, at most, a 10-qubit system \cite{bradley2019ten}.\footnote{This isn't necessarily a fundamental limit; we are simply stating a current practical limitation with NV centers.} If a QCS protocol needed 20 qubits, then it would unfortunately be incompatible with implementation with a single NV center. As an example of quantum-sensing hardware that has limited controllability, gases of cold atoms can be entangled and offer state-of-the-art sensing performance \cite{pezze2018quantum}, but don't typically allow for single-atom addressing or for universal quantum circuits---which the QCS protocols presented in this Perspective typically assume. In general, the simpler a QCS protocol is---in terms of number of qubits or qumodes required, the depth of the circuit, and the requirements on how sophisticated the control of the physical system needs to be---the more likely it is that it will be possible to find a hardware platform that has state-of-the-art sensitivity relevant to the target application and that can also successfully run the QCS protocol. However, more complicated QCS protocols that can in principle offer larger advantages, while probably not the first to achieve practical advantage, may ultimately inspire the development of sensing hardware that can successfully run them.

The prospects for practical advantage with simpler QCS protocols appear quite promising. While a practical advantage in quantum \emph{computing} requires that the computation being performed be intractable classically---which means it needs to, at least, involve $\gtrsim$50 qubits and a fairly deep circuit---the requirements for a practical QCSA are much less stringent. Crucially, because QCSA is not about computing time but about sensing time, it is irrelevant how easy or difficult it is to simulate the QCS protocol on a classical computer---and we have covered several QCS protocols that use very small quantum systems (e.g., just a single qubit, or a handful of qubits, or a single qubit and qumode) and yet can achieve an advantage. Some of these small protocols might very well enable practical advantages using hardware platforms that exist today---such as single NV centers and superconducting cavity-qubit systems, which are both state-of-the-art conventional quantum sensors that outperform all other existing sensors (quantum and classical) for certain applications, so are ripe for conversion into quantum computational sensors for tasks associated with those applications.

The synthesis of quantum computation with quantum sensing that has recently emerged and that we have described as \emph{quantum computational sensing} is a subtle and rich topic for study. It explores how to define and reach the fundamental quantum limits of task-specific information extraction with and from quantum sensors, and enables a new kind of advantage from quantum technology. Quantum computational sensors expand the design space for quantum-enhanced sensing, opening up the possibility for practical quantum-sensing advantages that are not merely incremental improvements, but might even, for some applications, approach the revolutionary potential of the exponential speedups possible in quantum computing. 

\section*{Author contributions}

All authors contributed to surveying the literature and writing the manuscript. S.A.K. prepared the figures, and S.P. prepared the tables. P.L.M. supervised the effort.

\section*{Acknowledgements}

We would like to thank Johannes~Borregaard, Vedran~Dunjko, Valla~Fatemi, James~Gardner, Emil~Khabiboulline, Vladimir~Kremenetski, Jérémie~Laydevant, Shi-Yuan~Ma, Benjamin~Malia, Aleksandr~Mokeev, Erich~Mueller, Tatsuhiro~Onodera, Mathieu~Ouellet, Mandar~Sohoni, Hakan~T\"ureci, Nathan~Wiebe, Fan~Wu and Ryotatsu~Yanagimoto for helpful discussions.

We gratefully acknowledge financial support from the Air Force Office of Scientific Research under award number FA9550-22-1-0203. We thank NTT Research for their financial and technical support. P.L.M. acknowledges membership in the CIFAR Quantum Information Science Program as an Azrieli Global Scholar.

\bibliographystyle{mcmahonlab}
\bibliography{references}

\appendix

\section*{Appendix} 







\section{Connecting stochastic displacement sensing to sensing of thermal states}

In this appendix section we discuss how the weak-light-detection scenario considered in quantum-enhanced optical imaging schemes discussed in e.g. Refs.~\cite{khabiboulline_quantum-assisted_2019, khabiboulline_optical_2019,mokeev_enhancing_2025, gardner_quantum_2026} have connections to \textit{stochastic} displacement sensing.

Consider $M$ bosonic receivers configured to receive light from a far away source. From Ref.~\cite{tsang_quantum_2011}, Eq.~(B1),~(B2) or Ref.~\cite{khabiboulline_optical_2019}, supplemental material Eq.~(5), the initial density matrix defining the state of received light at a quantum imaging setup can be defined as: 
\begin{align}
    \rho = \int d\vec{\alpha}~P(\vec{\alpha}) \ket{\vec{\alpha}}\bra{\vec{\alpha}}
    \label{eq:rhotherm}
\end{align}
where $\vec{\alpha} = (\alpha_1,\alpha_2,\ldots,\alpha_M)$ for $M$ receiver modes, and where the probability distribution $P(\vec{\alpha})$ is given by:
\begin{align}
    P(\vec{\alpha}) = \frac{1}{\pi^2\det \bm{\Gamma}}~\exp \left[ -
        \vec{\alpha}^{*}~
    \bm{\Gamma}^{-1}~\vec{\alpha}^T
    \right].
    \label{eq:pdfMM}
\end{align}
where $\bm{\Gamma}$ is the covariance matrix. Formally, the matrix elements of the covariance matrix are defined via:
\begin{align}
    [\bm{\Gamma}]_{jk} \equiv \Gamma_{jk} = \int d\vec{\alpha}~P(\vec{\alpha})\alpha_j^*\alpha_k  \equiv \mathbb{E}_{\vec{\alpha}}[\alpha_j^*\alpha_k]
\end{align}
where the final expression is used to define $\mathbb{E}_{\vec{\alpha}}[\cdot]$.

Now we can see how Eq.~(\ref{eq:rhotherm}) can also be viewed from the perspective of {stochastic displacement sensing}. In particular, suppose that we again consider $M$ bosonic receivers, initialized in the multimode vacuum state $\ket{\vec{0}}$, which in a given sensing instance are acted on by a multimode displacement operator $D(\vec{\alpha})$ (in response to a signal). Then \textit{for one sensing instance}, the initial state of the receiver is the pure state given by $\ket{\psi(\vec{\alpha})} = D(\vec{\alpha})\ket{\vec{0}}$. 

Next, we consider the scenario where every new sensing instance leads to a stochastic displacement $D(\vec{\alpha})$, according to some probability distribution $P(\vec{\alpha})$. In this case, the receiver state will be a mixed state defined by the density matrix:
\begin{align}
    \rho = \int d\vec{\alpha} ~P(\vec{\alpha})\ket{\psi(\vec{\alpha})}\bra{\psi(\vec{\alpha})} =  \int d\vec{\alpha} ~P(\vec{\alpha})D(\vec{\alpha})\ket{\vec{0}}\bra{\vec{0}}D^{\dagger}(\vec{\alpha}) = \int d\vec{\alpha} ~P(\vec{\alpha})\ket{\vec{\alpha}}\bra{\vec{\alpha}}
\end{align}
which is of course identical to Eq.~(\ref{eq:rhotherm}). It is in this sense that we consider the thermal state sensing problem as being equivalent to the problem of sensing stochastic displacements, when $P(\vec{\alpha})$ is given by Eq.~(\ref{eq:pdfMM}). Furthermore, we note that for the thermal case, displacements are on average zero, namely $\int d\vec{\alpha}~ P(\vec{\alpha})\alpha_i = 0$. All relevant information is carried by the covariances of the stochastic displacements, given by the covariance matrix $\bm{\Gamma}$.

We finally make a couple of observations. We have discussed here the connection between weak thermal light imaging and stochastic displacement sensing in the limit of many sensing instances, which yields the same density matrix for both cases. However, one could imagine that individual sensing instances could be modeled differently between the two cases, without changing the density matrix. This is ultimately a question that depends on the specific sensing apparatus. Secondly, one could consider the sensing of stochastic displacement channels defined by arbitrary $P(\vec{\alpha})$, which motivates an analysis of this problem beyond the case of thermal light sensing.

\begin{center}
    \rule{30mm}{1pt}
\end{center}

\end{document}